\documentclass[aps,prr,twocolumn,longbibliography]{revtex4-2}
\usepackage[T1]{fontenc}
\usepackage[latin9]{inputenc}
\usepackage{verbatim}
\usepackage{float}
\usepackage{amsmath}
\usepackage{amssymb}
\usepackage{graphicx}
\usepackage{caption}
\usepackage{placeins}
\usepackage{capt-of}
\usepackage{ragged2e}
\usepackage{tikz}
\usetikzlibrary{backgrounds}
\usepackage[justification=justified,singlelinecheck=false]{caption}

\makeatletter

\usepackage{dblfloatfix}
\usepackage{float}
\usepackage{afterpage}
\ifdefined\showcaptionsetup
 \PassOptionsToPackage{caption=false}{subfig}
\fi
\usepackage{subfig}
\makeatother

\usepackage{babel}
\begin{document}
\title{Weakly nonlinear dynamics of follower-force-induced oscillations in Cosserat rods}
\author{Mohamed Warda}
\email{mrmaw2@cam.ac.uk}

\affiliation{Cavendish Laboratory, University of Cambridge, J.J. Thomson Avenue,
Cambridge CB3 0US, United Kingdom}
\affiliation{Department of Applied Mathematics and Theoretical Physics, Centre
for Mathematical Sciences, University of Cambridge, Cambridge CB3
0WA, United Kingdom}
\begin{abstract}

We investigate the weakly nonlinear dynamics of the flutter instability of a planar Cosserat rod in a viscous fluid driven by a terminal follower force. Building on our previous linear stability analysis, we derive Stuart-Landau amplitude equations governing the emergence and saturation of oscillations near the Hopf bifurcation. The Landau coefficients are expressed in terms of the critical eigenmode, its adjoint, and quadratic correction fields arising from the asymptotic expansion. The reduced theory predicts the criticality of the Hopf bifurcations, the saturated oscillation amplitude, and the nonlinear frequency correction, in quantitative agreement with direct numerical simulations for representative material parameter regimes. Applying the theory to both stability boundaries of the finite oscillatory window, we find that the lower Hopf bifurcations are supercritical, whereas the upper restabilization threshold is subcritical. These results provide a normal-form description of follower-force-driven oscillations in geometrically exact Cosserat rods and a reduced framework for studying nonconservative elastohydrodynamic instabilities in soft robotic structures and active filaments.
\end{abstract}
\maketitle

\section{Introduction}

Soft slender structures interacting with viscous fluids arise in a broad range of biological and engineering settings, including active cytoskeletal filaments, synthetic micro-swimmers, underwater soft robotic manipulators, and minimally invasive continuum devices \citep{Laskar2013, DeCanio2017, Chirikjian1994, Ranzani2016, Campisano2020,Boyer2021, Gazzola2025}. In such systems the Reynolds number is often small, so inertia is negligible and the dynamics are governed primarily by the balance between elasticity and viscous dissipation. Nevertheless, nonconservative forcing can continuously inject energy into the structure and generate self-excited oscillations, enabling autonomous beating, propulsion, pumping, and mixing \citep{Laskar2013, DeCanio2017, Laskar2017, Clarke2024, Warda2025, Schnitzer2025}. The underlying nonconservative instability mechanism has its origins in classical studies of elastic stability \citep{Beck1952, Bolotin1963, Ziegler1968}.

A canonical mechanism for such behavior is the follower force instability \citep{Beck1952, Bolotin1963, Ziegler1968}, in which a compressive load remains tangent to the deformed configuration of a rod or filament. Because the load direction depends on the instantaneous configuration of the rod, the resulting linearized dynamics is non-self-adjoint. Unlike conservative Euler buckling, such systems can lose stability through a Hopf bifurcation even in the overdamped limit, as demonstrated in active filament and follower force models \citep{Laskar2013, Laskar2017, DeCanio2017, Clarke2024, Warda2025, Schnitzer2025}. Since the classical studies of Beck's column and related nonconservative elastic systems \citep{Beck1952,Bolotin1963,Ziegler1968,Carr1979,Chen1987,Koch2000,Wang2004}, follower force models have re-emerged in soft matter and active filament contexts, where they provide effective descriptions of tip-directed motor forces or localized active stresses \citep{Laskar2013, DeCanio2017, Clarke2024, Link2024, Warda2025, Schnitzer2025}. Closely related active filament models driven by distributed tangential body forces have long been studied as models of molecular-motor-driven filaments and motility assays, beginning with the pioneering work of Sekimoto \citep{Sekimoto1995} and many subsequent developments \citep{Holder2015, Ling2018, Fily2020}. Recently, models of active filaments utilizing strain-dependent force and torque densities have been found to exhibit complex shape space dynamics \citep{Nemeth2016}. The follower force model considered here may be viewed as a boundary-driven counterpart of these distributed active force descriptions.

While linear stability theory determines the threshold of oscillatory instability, it does not describe the finite-amplitude state selected beyond onset. Near the threshold, weakly nonlinear analysis provides the natural asymptotic framework for deriving reduced amplitude equations that capture the slow modulation and saturation of unstable modes. Such methods provide a universal framework for describing pattern-forming and oscillatory instabilities across fluid mechanics, active matter, and elasticity \citep{Cross1993, Kuznetsov2023, Barkley2006, Tchoufag2015, Schnitzer2025}. In particular, Stuart-Landau equations determine whether a Hopf bifurcation is supercritical or subcritical and predict amplitude scaling laws and nonlinear frequency shifts, as described in the classical theory of Hopf bifurcations and amplitude equations \citep{Cross1993}.

These ideas have recently been applied to the follower force model for overdamped filaments. Schnitzer \citep{Schnitzer2025} analyzed the three-dimensional follower force model for inextensible, unshearable filaments near the onset of instability, showing that the transition is governed by a symmetry-induced double Hopf bifurcation. The resulting reduced amplitude equations generalize the Stuart-Landau normal form and explain the emergence and selection of non-planar whirling states from planar modes. Related nonlinear studies of active or axially driven filaments have also demonstrated spontaneous oscillation, morphological transitions, and self-organized beating states
\citep{Laskar2013, DeCanio2017, Link2024, Warda2025}.

The present work addresses the weakly nonlinear dynamics of the geometrically exact Cosserat-rod model introduced in our preceding study of pressure-driven soft robotic arms in viscous fluids \citep{Warda2025}. That work established the existence of a finite window of oscillatory instability bounded by Hopf bifurcations and demonstrated stable limit-cycle oscillations in the fully nonlinear dynamics. The present paper develops an analytical weakly nonlinear theory for these transitions. By performing a multiple-scale expansion about the compressed straight state and enforcing cubic-order solvability using the adjoint eigenmode of the non-Hermitian linear operator, we derive a family of Stuart-Landau amplitude equations parameterized by the constitutive properties of the rod.

Beyond the analytical derivation, the resulting amplitude equations are evaluated for several representative material parameter sets, allowing systematic comparison between analytical predictions and direct nonlinear simulations. The theory predicts whether each Hopf bifurcation is supercritical or subcritical, together with the saturated oscillation amplitude and nonlinear frequency correction. These predictions are compared directly with numerical simulations of the full Cosserat-rod equations. We furthermore examine both the onset and restabilization boundaries of the finite instability window established in \citep{Warda2025}, demonstrating how the weakly nonlinear theory distinguishes supercritical and subcritical behavior at the two stability thresholds.



Unlike the three-dimensional inextensible Kirchhoff-filament formulation considered by Schnitzer \citep{Schnitzer2025}, the present work treats geometrically exact planar Cosserat rods with stretch and shear. These additional degrees of freedom introduce longitudinal correction fields and quadratic couplings absent from the Kirchhoff formulation, modifying both the asymptotic hierarchy and the resulting Landau coefficients. Moreover, the finite instability window identified previously for the Cosserat model enables analysis of both stability boundaries.

The remainder of the paper is organized as follows. Section \ref{sec:problemoverview} summarizes the Cosserat-rod model and governing equations. Section \ref{sec:hopfbifurcation} reviews the linear Hopf instability and near-threshold scaling. Section \ref{sec:weaklynonlinearreduction} develops the multiple-scale expansion, while Sec. \ref{sec:stuartlandau} derives the Stuart-Landau amplitude equation. Section \ref{sec:simulations} compares the resulting weakly nonlinear predictions with nonlinear simulations across several constitutive parameter regimes and examines the nonlinear behavior at both stability boundaries of the finite instability window. Finally, Sec. VII concludes with a discussion of the implications and possible extensions.



\section{Problem Overview \label{sec:problemoverview}}
We briefly review the notation and governing equations of the Cosserat-rod model introduced in \citep{Warda2025}. The configuration of a planar Cosserat rod at time $t$ is described by its
centerline $\boldsymbol{r}(u,t)$ and orthonormal frame vectors
$\boldsymbol{e}_{1}(u,t),\boldsymbol{e}_{2}(u,t)$,
rigidly attached to the cross-section at material parameter $u$. Relative to a fixed frame, we may introduce coordinates $(x(u, t), y(u,t), \theta(u,t))$ such that
\begin{equation}
    \begin{aligned}
        \boldsymbol{r}(u,t) &= (x(u,t), y(u,t))\\
        \boldsymbol{e}_{1}(u,t) &= (\cos\theta(u,t), \sin\theta(u,t))\\
        \boldsymbol{e}_{2}(u,t) &= (-\sin\theta(u,t), \cos\theta(u,t))
    \end{aligned}
\end{equation}
We express the equations of motion of the rod in the moving frame $(\boldsymbol{e}_{1}, \boldsymbol{e}_{2})$ in terms of velocities $(v_{1},v_{2},\Omega)$ and strains $(h_{1},h_{2},\Pi)$, related to the spatial and temporal derivatives of the centerline and frame vectors by
\begin{align}
v_{1} & =\boldsymbol{e}_{1}\cdot\partial_{t}\boldsymbol{r},\quad v_{2}=\boldsymbol{e}_{2}\cdot\partial_{t}\boldsymbol{r},\quad\Omega=\boldsymbol{e}_{2}\cdot\partial_{t}\boldsymbol{e}_{1},\nonumber \\
h_{1} & =\boldsymbol{e}_{1}\cdot\partial_{u}\boldsymbol{r},\quad h_{2}=\boldsymbol{e}_{2}\cdot\partial_{u}\boldsymbol{r},\quad\Pi=\boldsymbol{e}_{2}\cdot\partial_{u}\boldsymbol{e}_{1}.
\label{eq:definitions}
\end{align}
The quantities $(h_1,h_2,\Pi)$ characterize the local deformation of the rod. Here, $h_1$ and $h_2$ are the axial and shear strain components, respectively, while $\Pi$ is the bending strain (curvature).
Their kinematic counterparts $(v_1,v_2,\Omega)$ are the translational and angular velocities of the cross section expressed in the moving frame $(\boldsymbol{e} _1,\boldsymbol{e}_2)$ \citep{Antman2005}. 

In the absence of inertia, we may write down the dynamics of the Cosserat
rod covariantly in terms of stresses $(F_{1},F_{2},M)$ and moving frame components of the body force and body torque densities $(f_{1},f_{2},m)$. In this overdamped limit, the force and torque balance equations may be written as
\begin{equation}
    \begin{aligned}
        \partial_u F_1 - \Pi F_2 + f_1 &= 0\\
        \partial_u F_2 + \Pi F_1 + f_2 &= 0\\
        \partial_u M + h_1 F_2 - h_2 F_1 + m &= 0
    \end{aligned}\label{eq:explicitdynamics}
\end{equation}
The corresponding internal stress resultants are the force components $(F_1,F_2)$ and bending moment $M$. The quantities $(f_1,f_2)$ and $m$ denote external body force and body torque densities defined per unit material coordinate $u$ (reference length). We note that the balance laws are written per unit reference length $u$, consistent with the geometrically exact Cosserat formulation \citep{Antman2005}.

We consider a Cosserat rod of rest length
$L$ and a stress-free configuration 
\begin{equation}
\bar{\boldsymbol{r}}(u)=(u,0),\quad\bar{\boldsymbol{e}}_{1}(u)=(1,0),\quad\bar{\boldsymbol{e}}_{2}(u)=(0,1).
\end{equation}
The components of the strain in this configuration are $(\bar{h}_1, \bar{h}_2, \bar{\Pi})=(1,0,0)$. We consider a rod subject to linear Stokes drag
\begin{equation}
    f_1 = -\gamma_1 v_1, \quad f_2 = -\gamma_2 v_2, \quad m = -\gamma_3 \Omega,
    \label{eq:drag}
\end{equation}
and assume a quadratic elastic energy 
\begin{equation}
    \mathcal{U}[\varphi] = \frac{1}{2}\int_{0}^{L}\mathrm{d}u\,k_{1}(h_{1}-1)^{2}+k_{2}h_{2}^{2}+k_{3}\Pi^{2}
\end{equation}
such that
\begin{equation}
    F_1 = k_1 (h_1 - 1), \quad F_2 = k_2 h_2, \quad M = k_3 \Pi.
    \label{eq:constitutive}
\end{equation}
We note that a critical approximation is made in the linear Stokes drag law (\ref{eq:drag}). The constitutive relation $(f_1, f_2, m) = \Gamma (v_1, v_2, \Omega)$ with constant $\Gamma$ is intended as a local phenomenological resistance law defined per unit reference coordinate $u$. It should not be interpreted as the exact pull-back of a constant-coefficient resistive-force law defined per unit current arclength $s$, for which the referential resistance tensor would require a metric factor $\sqrt{h_1^2 + h_2^2}$. More complete hydrodynamic descriptions of extensible and shearable Cosserat rods derived from slender-body theory lead to strain-dependent and generally nonlocal mobility relations, as demonstrated in \citep{Garg2023}. Our constant referential resistance tensor is instead adopted as the simplest local, positive-definite constitutive closure consistent with the material-coordinate formulation, allowing an analytically tractable weakly nonlinear theory consistent with the model presented in \citep{Warda2025}. Consequently, incorporating the metric dependence of the hydrodynamic
resistance would in general modify the nonlinear terms and hence the
numerical simulations as well as the values of the Landau coefficients derived below; the present
coefficients should therefore be understood as those associated with the
constant referential resistance model (\ref{eq:drag}), and consequently those associated with our previous work in \citep{Warda2025}. A systematic
extension of the weakly nonlinear analysis to more accurate hydrodynamic resistance laws is beyond the scope of the present work.

Together with the kinematic relations (\ref{eq:definitions}), the constitutive laws (\ref{eq:constitutive}) and (\ref{eq:drag}) and the balance equations (\ref{eq:explicitdynamics})define a closed system of three nonlinear partial differential equations for $(x, y, \theta)$. We assume that the rod is clamped at $u = 0$
\begin{equation}
    x(0, t) = 0,\quad y(0, t) = 0, \quad \theta(0, t) = 0
    \label{eq:clamping}
\end{equation}
and subject to a follower force $\mathcal{F}$ at $u = L$
\begin{equation}
    F_1(L, t) = -\mathcal{F},\quad F_2(L, t) = 0, \quad M(L, t) = 0.
    \label{eq:follower}
\end{equation}
The full nonlinear problem is specified by the time-dependent partial differential equations for $(x, y, \theta)$, the boundary conditions in Eq. (\ref{eq:clamping}) and Eq. (\ref{eq:follower}), and an appropriate initial condition on $(x, y, \theta)$. Details of the simulation of this nonlinear system of PDEs may be found in \cite{Warda2025}. As shown in \cite{Warda2025}, the nonlinear system possesses a trivial time-independent solution representing a compressed but otherwise undeformed rod
\begin{equation}
x^{(0)}=\nu u,\quad y^{(0)}=0,\quad\theta^{(0)}=0,
\label{eq:basestate}
\end{equation}
where $\nu=1-\mathcal{F}/k_{1}$ is a compression induced by the applied follower force. This compressed equilibrium serves as the base state for the linear stability analysis in Sec. \ref{sec:hopfbifurcation} and the subsequent weakly nonlinear analysis.

\section{Linear Stability Analysis \label{sec:hopfbifurcation}}
We briefly summarize the linear stability analysis established in \citep{Warda2025}, emphasizing the results required for the weakly nonlinear analysis developed below. We non-dimensionalize the problem with respect to the length scale $L$ and the time scale $\gamma_2L^4/k_3$, with respect to which we express our equations in terms of the dimensionless constitutive parameters
\begin{equation}
    \tilde{k}_1 = \frac{k_1L^2}{k_3},\quad \tilde{k}_2 = \frac{k_2L^2}{k_3},\quad \tilde{\gamma}_1 = \frac{\gamma_1}{\gamma_2}, \quad \tilde{\gamma}_3 = \frac{\gamma_3}{\gamma_2 L^2}
    \label{eq:dimensionlessconstitutive}
\end{equation}
and the dimensionless control parameter
\begin{equation}
    \tilde{\mathcal{F}} = \frac{\mathcal{F}L^2}{k_3}
    \label{eq:dimensionlesscontrol}
\end{equation}
The interpretations of (\ref{eq:dimensionlessconstitutive}) and (\ref{eq:dimensionlesscontrol}) are summarized in Table \ref{tab:summary}. Throughout this work, for simplicity, we fix the translational drag ratio $\tilde{\gamma}_1 = 1/2$, consistent with the classical local resistive-force approximation for slender filaments in Stokes flow, for which the transverse drag is approximately twice the tangential drag \citep{Lighthill1976}.
\begin{table}[h]
  \centering
  \begin{ruledtabular}
    \begin{tabular}{ccc}
      Symbol & Definition & Interpretation  \\
      \colrule
      $\tilde{k}_{1}=k_{1}L^{2}/k_{3}$ & Stretch stiffness ratio       & Compressibility       \\
      $\tilde{k}_{2}=k_{2}L^{2}/k_{3}$ & Shear stiffness ratio        & Shear rigidity        \\
      $\tilde{\gamma}_{1}=\gamma_{1}/{\gamma_{2}}$ & Translational drag ratio& Drag anisotropy       \\
      $\tilde{\gamma}_{3}=\gamma_{3}/{\gamma_{2}L^{2}}$ & Rotational drag ratio& Angular dissipation       \\
      $\tilde{\mathcal{F}}=\mathcal{F}L^{2}/{k_{3}}$ & Follower force strength        & Control parameter     \\
    \end{tabular}
  \end{ruledtabular}
 \caption{\parbox{\linewidth}{\justifying\noindent \newline A summary of the dimensionless parameters $(\tilde{k}_{1},\tilde{k}_{2}, \tilde{\gamma}_{1}, \tilde{\gamma}_{3}, \tilde{\mathcal{F}})$ introduced in this section and their interpretations.}\label{tab:summary}}
\end{table}

Linearization of the equations of motion of the rod about the compressed straight equilibrium (\ref{eq:basestate}) leads to the system
\begin{equation}
    \partial_t \xi_\alpha = \mathcal{L}_{\alpha\beta} \xi_\beta
    \label{eq:linearization}
\end{equation}
where $\xi_\alpha = (x^{(1)}, y^{(1)}, \theta^{(1)})$ is a vector of perturbations of the fields $(x, y, \theta)$ and 
\begin{equation}
    \mathcal{L}_{\alpha\beta} = \begin{bmatrix}\tilde{\gamma}_1^{-1}\tilde{k}_{1}\partial_{u}^{2} & 0 & 0\\
0 & \tilde{k}_{2}\partial_{u}^{2} & -\mu\partial_{u}\\
0 & \tilde{\gamma}_3^{-1}\mu\partial_{u} & \tilde{\gamma}_3^{-1} \left(\partial_{u}^{2}-\nu\mu\right)
\end{bmatrix}
\end{equation}
is a linear differential operator with the associated boundary conditions
\begin{equation}
    x^{(1)} = y^{(1)} = \theta^{(1)} = 0 \quad \mathrm{at}\,\, u = 0
\end{equation}
and 
\begin{equation}
    \partial_ux^{(1)} = \partial_uy^{(1)} - \nu \theta^{(1)} = \partial_u\theta^{(1)} = 0 \quad \mathrm{at}\,\, u = 1.
\end{equation}
We define the quantity $\mu \equiv \tilde{k}_2\nu + \mathcal{F}$ for convenience. The operator $\mathcal{L}$ is generically non-Hermitian for $\tilde{\mathcal{F}} \neq 0$ with respect to the canonical $\Gamma$-inner product
\begin{equation}
    (\Psi, \Phi)_\Gamma = \int_0^1 \mathrm{d}u \, \Psi^\dagger(u) \Gamma \Phi(u),
    \label{eq:inner product}
\end{equation}
where $\dagger$ denotes the standard conjugate transpose and $\Gamma = \mathrm{diag}(\tilde{\gamma}_1, 1, \tilde{\gamma}_3)$ is a constant diagonal friction tensor.
\begin{figure}[t]
\RaggedRight
\includegraphics[scale=0.55]{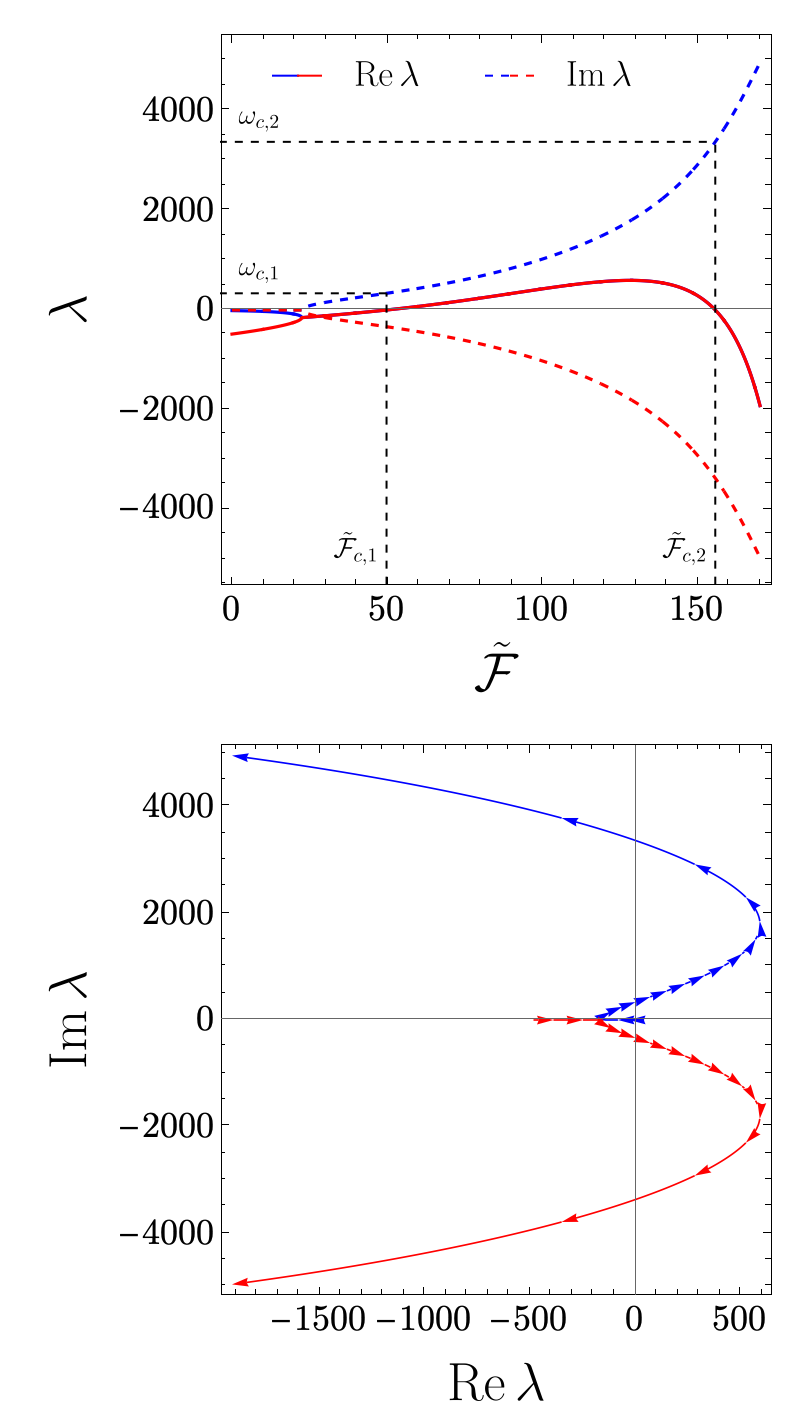}
\caption{\parbox{\linewidth}{\justifying\noindent \newline \label{fig:hopf}Reentrant linear stability spectrum of the planar Cosserat rod for
$\tilde{k}_{1}=200$, $\tilde{k}_2 = 10^4$, and
$\tilde{\gamma}_{3}=10^{-4}$.
\textbf{Top:}
Real (solid) and imaginary (dashed) parts of the leading eigenvalues
$\lambda$ as functions of the nondimensional follower force
$\tilde{\mathcal{F}}$.
The dashed black lines indicate the lower and upper stability thresholds,
$\tilde{\mathcal{F}}_{c,1}$ and
$\tilde{\mathcal{F}}_{c,2}$,
together with the corresponding Hopf frequencies
$\omega_{c,1}$ and
$\omega_{c,2}$.
The straight equilibrium is linearly stable outside this interval and
unstable within it, giving rise to a finite window of oscillatory
instability.
\textbf{Bottom:}
Trajectories of the corresponding leading eigenvalues in the complex
plane as the follower force is varied.
The arrows indicate the direction of increasing
$\tilde{\mathcal{F}}$,
illustrating the motion of the eigenvalues through the two Hopf
bifurcations that bound the oscillatory window.}}
\end{figure}

\begin{figure}[t]
\centering
\includegraphics[scale=0.6]{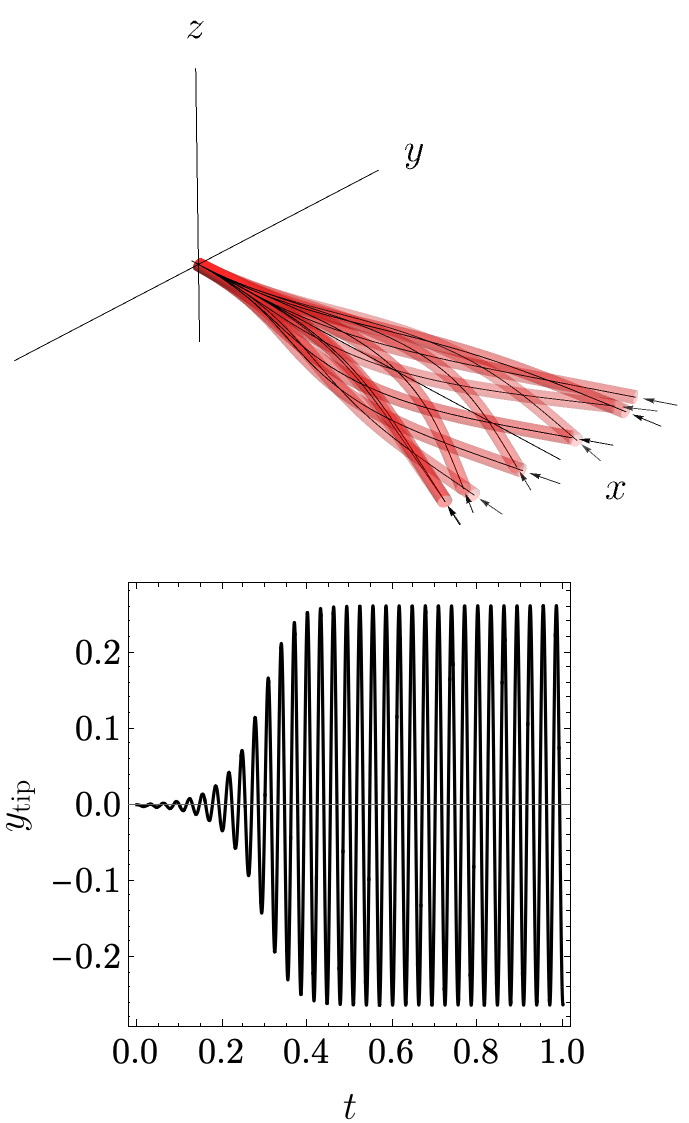}

\caption{\parbox{\linewidth}{\justifying\noindent \newline \label{fig:nonlinear} Emergence of limit cycle oscillations of the rod beyond the instability threshold, visualized through snapshots (top) and of the tip of the rod (bottom) for $\tilde{k}_1 = \tilde{k}_2 = 10^4$, $\tilde{\gamma}_1 = 1/2$, $\tilde{\gamma}_3 = 10^{-4}$, and $\tilde{\mathcal{F}} = 40$ (beyond the stability threshold $\tilde{\mathcal{F}}_c \approx 37.66$). The arrows indicate the configuration-dependent direction of the follower force. Numerical simulations were carried out using the methodology in \citep{Warda2025}.}}
\end{figure}

As shown in \citep{Warda2025}, the spectrum of the operator $\mathcal{L}$ undergoes Hopf bifurcations as the dimensionless follower force $\tilde{\mathcal{F}}$ is varied for a broad range of regimes of the parameters (\ref{eq:dimensionlessconstitutive}). A key result in \cite{Warda2025} is that, for sufficiently compressible Cosserat rods, the leading eigenvalue undergoes two Hopf bifurcations as the follower force increases, producing a finite interval of oscillatory instability bounded by a lower threshold $\tilde{\mathcal{F}}_{c,1}$ and an upper threshold $\tilde{\mathcal{F}}_{c,2}$. This reentrant behavior is absent from inextensible, unshearable filament models and provides the central motivation for extending the weakly nonlinear analysis beyond the onset bifurcation. An example of this is illustrated in Fig. \ref{fig:hopf}. Although Fig. \ref{fig:hopf} illustrates the spectrum for one representative parameter set, the weakly nonlinear theory developed below is formulated generally and will later be evaluated for several representative parameter regimes.

At a given stability threshold corresponding to a critical follower force $\tilde{\mathcal{F}}_c$ and a corresponding Hopf frequency $\omega_c$, we may write a planar beating solution of (\ref{eq:linearization}) as
\begin{equation}
    \begin{aligned}
        x^{(1)}(u, t) &= 0\\
        y^{(1)}(u, t) &= \tilde{A} Y(u) e^{i\omega_ct} + \mathrm{c.c.}\\
        \theta^{(1)}(u, t) &= \tilde{A} \Theta(u) e^{i\omega_ct} + \mathrm{c.c.}
    \end{aligned}
    \label{eq:neutral}
\end{equation}
where $\tilde{A}$ is a complex amplitude of small magnitude determined by the initial conditions. Normalizing the transverse eigenmode $(0, Y(u), \Theta(u))$ such that $Y(1) = 1$, we may write down the $y-$coordinate of the trajectory of the tip of the rod as
\begin{equation}
    y^{(1)}(1, t) = \tilde{A}  e^{i\omega_ct} + \mathrm{c.c.}
    \label{eq:tiplinear}
\end{equation}

While the nonlinear simulations of \citep{Warda2025} demonstrated saturation into stable limit-cycle oscillations, as shown in Fig. \ref{fig:nonlinear}, no analytical weakly nonlinear theory was developed there. The purpose of the present work is to derive the corresponding family of Stuart-Landau equations for various parameter regimes, determine the criticality of both the conventional lower Hopf bifurcation as well as the upper Hopf bifurcation observed in reentrant systems, and obtain analytical predictions for the saturated oscillation amplitude and nonlinear frequency correction near a given stability threshold.


\section{Weakly Nonlinear Analysis \label{sec:weaklynonlinearreduction}}
To determine the nonlinear saturation of the flutter instability, we consider values of the follower force close to the Hopf threshold and perform a weakly nonlinear analysis using the method of multiple scales \citep{Hinch1991}. We write the control parameter as
\begin{equation}
    \tilde{\mathcal{F}} = \tilde{\mathcal{F}}_c + \epsilon^2 \Delta
\end{equation}
where $0 < \epsilon \ll1$ is a small perturbation parameter and $\Delta = \mathcal{O}(1)$ measures the signed distance from threshold. 

The distinguished scaling above reflects the standard balance for a Hopf bifurcation: at the critical follower force $\tilde{\mathcal{F}} = \tilde{\mathcal{F}}_c$, the linearized dynamics admits neutrally stable oscillations at frequency $\omega_c$, whose leading-order contribution may be written as a critical eigenmode with complex amplitude $\tilde{A}$, as in (\ref{eq:neutral}). In the full nonlinear problem, cubic terms generically generate resonant forcing at the same frequency, which produces secular corrections growing in time like $t |\tilde{A}|^3$. Such corrections become comparable to the leading-order oscillation over long times $t = \mathcal{O}(|\tilde{A}|^{-2})$, indicating that nonlinear effects modulate the amplitude on a slow time scale. For $\tilde{\mathcal{F}}$ near $\tilde{\mathcal{F}}_c$, the real part of the leading eigenvalue $\lambda$ satisfies $\mathrm{Re}\, \lambda \sim \tilde{\mathcal{F}} - \tilde{\mathcal{F}}_c$, so that linear growth or decay occurs over the long time scale $\mathcal{O}(|\tilde{\mathcal{F}} - \tilde{\mathcal{F}}_c|^{-1})$. Balancing this linear time scale with the nonlinear modulation time scale yields the distinguished scaling $|\tilde{A}| = \mathcal{O}(|\tilde{\mathcal{F}} - \tilde{\mathcal{F}}_c|^{1/2})$, which underlies the multiple-scale expansion and the emergence of a Stuart-Landau amplitude equation.

Accordingly, we introduce a fast time $\tau = t$, resolving oscillations at the Hopf frequency $\omega_c$, together with a slow modulation time $T = \epsilon^2 t$. All rod variables are expanded in powers of $\epsilon$ about the compressed straight base state corresponding to $\tilde{\mathcal{F}} = \tilde{\mathcal{F}}_c$. Substitution into the full nonlinear Cosserat rod equations yields a hierarchy of boundary value problems at successive orders in $\epsilon$. Schematically, the hierarchy of boundary value problems at $n$th order takes the form
\begin{equation}
    \partial_\tau \xi^{(n)} = \mathcal{L} \xi^{(n)} + \Gamma^{-1} f^{(n)}
    \label{eq:hierarchy}
\end{equation}
along with the appropriate boundary conditions, where $\xi^{(n)}$ is a vector of $n$th order corrections of the fields $(x,y,\theta)$ associated with the multiple-scale expansion and $f^{(n)}$ is an $n$th order generalized forcing term composed of lower order contributions. The complete expansions and order-by-order equations and boundary conditions are given in Appendix \ref{sec:multiplescaleanalysis}.

The hierarchy (\ref{eq:hierarchy}) has a particularly simple structure. At leading order, one recovers the neutral Hopf eigenmode of the linearized problem. Because this mode is purely transverse, quadratic
nonlinearities cannot resonate with the critical frequency.
Instead, they generate only a static longitudinal correction and a
second-harmonic longitudinal response. Consequently, the
$\mathcal{O}(\epsilon^2)$ problem reduces to determining four
auxiliary fields describing the longitudinal displacement and
internal tension,
\begin{equation}
X_0(u), \qquad X_2(u), \qquad P_0(u), \qquad P_2(u),
\end{equation}
where the subscripts denote the zero-frequency and second-harmonic
components, respectively. Accordingly, the second-order corrections to $x$ and $F_1$ may be written as
\begin{equation}
    \begin{aligned}
        x^{(2)} &= A^2 X_2 e^{2i\omega_c\tau} + |A|^2 X_0 - \frac{\Delta}{2\tilde{k}_1} u + \mathrm{c.c.},\\
        F_1^{(2)} &= A^2 P_2 e^{2i\omega_c\tau} + |A|^2 P_0 - \frac{\Delta}{2} + \mathrm{c.c.}.
    \end{aligned}
\end{equation}
The fields $X_0$ and $X_2$ represent the mean and second-harmonic
longitudinal deformations induced by the transverse oscillation,
while $P_0$ and $P_2$ are the corresponding tension corrections. Although these fields do not modify the leading-order
oscillation directly, they provide the nonlinear feedback that
enters the resonant forcing at cubic order. The governing boundary-value problems of $X_0$ and $X_2$, derived in Appendix \ref{sec:weaklynonlinear}, are 
\begin{equation}
    \begin{aligned}
        \tilde{k}_1 X_0'' &= \Theta^* (i\omega_c \tilde{\gamma}_1 Y - \tilde{k}_1 Y'')\\
        &+ (\tilde{k}_2 - \tilde{k}_1) (Y' - \nu \Theta)^* \Theta'\\
        \tilde{k}_1 X_2'' &= 2 i \omega_c \tilde{\gamma}_1 X_2 + \Theta(i\omega_c \tilde{\gamma}_1 Y - \tilde{k}_1 Y'')\\
        &+ (\tilde{k}_2 - \tilde{k}_1) (Y' - \nu \Theta)\Theta'\\
    \end{aligned}
\end{equation}
with 
\begin{equation}
    \begin{aligned}
        &X_0 =  0, \quad X_2 = 0,\quad \mathrm{at} \,\, u = 0,\\
       & X_0' = -\frac{1}{2} \nu |\Theta|^2, \quad X_2' = -\frac{1}{2} \nu \Theta^2, \quad \mathrm{at} \,\,u=1.
    \end{aligned}
\end{equation}
The fields $P_0$ and $P_2$ are subsequently determined constitutively from the fields $X_0$ and $X_2$. 

At $\mathcal{O}(\epsilon^3)$, slow-time modulation, parameter
detuning, and the quadratic correction fields combine to generate
resonant transverse forcing terms at the critical Hopf frequency, taking the form
\begin{equation}
    \begin{aligned}
        f_2^{(3)} &= e^{i\omega_c \tau} \left(|A|^2A g_1 + \Delta A g_2 - Y \frac{\mathrm{d}A}{\mathrm{d}T} \right) + \mathrm{c.c.} + \mathrm{n.s.t.},\\
        f_3^{(3)} &= e^{i\omega_c \tau} \left(|A|^2A g_3 + \Delta A g_4 - \tilde{\gamma}_3\Theta \frac{\mathrm{d}A}{\mathrm{d}T} \right) + \mathrm{c.c. + \mathrm{n.s.t.}},\\
    \end{aligned}
    \label{eq:resonantforcing}
\end{equation}
where ``n.s.t.'' denotes non-secular terms (non-resonant contributions at
frequencies $0$, $2\omega_c$, $3\omega_c$, etc.) that do not enter the
solvability condition, and $g_i$ are functions of $u$ derived and presented explicitly in Appendix \ref{sec:weaklynonlinear}. These resonant terms must
be removed to prevent secular growth. Since the follower force
boundary condition renders the linearized operator
non-self-adjoint, the corresponding solvability condition \citep{Keener2018} is obtained by projecting the resonant forcing onto the adjoint critical eigenmode under the drag-weighted inner product (\ref{eq:inner product}).
This projection yields the Stuart-Landau amplitude equation
governing the slow evolution of the Hopf mode, presented in the
next section.

\section{The Stuart-Landau Equation \label{sec:stuartlandau}}
At leading order, the rod executes neutral oscillations at the Hopf frequency $\omega_c$ with spatial structure given by the critical eigenmode $(0, Y(u), \Theta(u))$. We therefore write the first order solution as 
\begin{equation}
    \begin{aligned}
        x^{(1)}(u, \tau, T) &= 0,\\
        y^{(1)}(u, \tau, T) &= A(T) Y(u) e^{i\omega_c\tau} + \mathrm{c.c.},\\
        \theta^{(1)}(u, \tau, T) &= A(T) \Theta(u) e^{i\omega_c\tau} + \mathrm{c.c.},
        \label{eq:solutionlinear}
    \end{aligned}
\end{equation}
where $A(T)$ is a complex amplitude varying only on the slow time scale.

The follower force boundary condition renders the linearized operator non-self-adjoint. Consequently, the cubic-order solvability condition must be imposed using the critical adjoint eigenmode $(0, \Psi_Y(u), \Psi_\Theta(u))$ under the drag-weighted inner product (\ref{eq:inner product}). Carrying out this projection (Appendix \ref{sec:weaklynonlinear}) removes resonant forcing at frequency $\omega_c$ and yields the amplitude equation
\begin{equation}
    \frac{\mathrm{dA}}{\mathrm{d}T} = \alpha |A|^2A + \beta \Delta A.
    \label{eq:SL}
\end{equation}
This  is the Stuart-Landau normal form governing the onset of beating in the planar Cosserat rod. The complex Landau coefficients $\alpha$ and $\beta$, derived explicitly in Appendix \ref{sec:weaklynonlinear}, are given by
\begin{equation}
    \begin{aligned}
        \alpha &= \frac{\displaystyle\int_0^1 \mathrm{d}u \, \Psi_Y^* g_1 + \Psi_\Theta^* g_3}{\displaystyle\int_0^1 \mathrm{d}u \, \Psi_Y^* Y + \tilde{\gamma}_3\Psi_\Theta^* \Theta},\\
        \beta &= \frac{\displaystyle\int_0^1 \mathrm{d}u \, \Psi_Y^* g_2 + \Psi_\Theta^* g_4}{\displaystyle\int_0^1 \mathrm{d}u \, \Psi_Y^* Y + \tilde{\gamma}_3\Psi_\Theta^* \Theta},
    \end{aligned}
    \label{eq:alphabetaformulas}
\end{equation}
The Landau coefficients $\alpha$ and $\beta$ depend parametrically on the constitutive properties of the Cosserat rod through the critical eigenmode, the corresponding adjoint eigenmode, and the quadratic correction fields computed in Appendix \ref{sec:weaklynonlinear}. 

Equation (\ref{eq:SL}) constitutes the reduced normal-form
description of the follower force instability. Its real
coefficients determine the evolution and saturation of the
oscillation amplitude, while its imaginary coefficients
determine the slow phase evolution and hence the nonlinear
frequency correction. To make this separation explicit, we write
\begin{equation}
A(T)=R(T)e^{i\varphi(T)},
\end{equation}
where $R(T)\geq0$. Substitution into (\ref{eq:SL}) gives
\begin{align}
\frac{\mathrm{d}R}{\mathrm{d}T}
&=
\left[
\operatorname{Re}(\beta)\Delta
+
\operatorname{Re}(\alpha)R^2
\right]R,
\label{eq:radial_amplitude}
\\
\frac{\mathrm{d}\varphi}{\mathrm{d}T}
&=
\operatorname{Im}(\beta)\Delta
+
\operatorname{Im}(\alpha)R^2.
\label{eq:phase_amplitude}
\end{align}
The trivial solution $R=0$ represents the straight state.
Linearizing Eq.~\eqref{eq:radial_amplitude} about $R=0$
shows that this state is unstable when
$\operatorname{Re}(\beta)\Delta>0$ and stable when
$\operatorname{Re}(\beta)\Delta<0$.

A nonzero constant-modulus solution satisfies
\begin{equation}
R_*^2
=
-\frac{\operatorname{Re}(\beta)}
{\operatorname{Re}(\alpha)}
\Delta,
\label{eq:saturated_amplitude}
\end{equation}
and therefore exists when
$\operatorname{Re}(\beta)\Delta$ and
$\operatorname{Re}(\alpha)$ have opposite signs.
Its radial stability is determined by
\begin{equation}
\left.
\frac{\mathrm{d}}{\mathrm{d}R}
\left\{
\left[
\operatorname{Re}(\beta)\Delta
+\operatorname{Re}(\alpha)R^2
\right]R
\right\}
\right|_{R=R_*}
=
-2\operatorname{Re}(\beta)\Delta.
\end{equation}
Thus, when $\operatorname{Re}(\alpha)<0$, the finite-amplitude
branch lies on the linearly unstable side of the threshold
and is stable, corresponding to a supercritical Hopf
bifurcation. When $\operatorname{Re}(\alpha)>0$, the cubic
branch lies on the linearly stable side and is unstable,
corresponding to a subcritical Hopf bifurcation. In the
subcritical case, the cubic normal form alone does not
determine the stable large-amplitude dynamics.

For the constant-modulus solution, the slow phase evolves as
$\varphi(T)=\sigma T+\phi$, where
\begin{equation}
\sigma
=
\left[
\operatorname{Im}(\beta)
-
\frac{
\operatorname{Im}(\alpha)\operatorname{Re}(\beta)
}{
\operatorname{Re}(\alpha)
}
\right]\Delta .
\label{eq:slow_frequency}
\end{equation}
Using the eigenmode normalization $Y(1)=1$, the leading-order
tip displacement is
\begin{equation}
y(1,t)
\simeq
2\epsilon R_*
\cos\left[
\left(\omega_c+\epsilon^2\sigma\right)t+\phi
\right].
\end{equation}
Since
$\tilde{\mathcal F}-\tilde{\mathcal F}_c
=\epsilon^2\Delta$, the physical amplitude and frequency
predictions may be written as
\begin{align}
A_{\mathrm{tip}}
&=
2\sqrt{
-\frac{\operatorname{Re}(\beta)}
{\operatorname{Re}(\alpha)}
\left(
\tilde{\mathcal F}
-\tilde{\mathcal F}_c
\right)
}, \label{eq:tipamplitude}
\\
\omega
&=
\omega_c+
\left[
\operatorname{Im}(\beta)
-
\frac{
\operatorname{Im}(\alpha)\operatorname{Re}(\beta)
}{
\operatorname{Re}(\alpha)
}
\right]
\left(
\tilde{\mathcal F}
-\tilde{\mathcal F}_c
\right).\label{eq:tipfrequency}
\end{align}
These expressions provide the weakly nonlinear predictions
tested against the full nonlinear simulations in Sec. \ref{sec:simulations}. Note that the amplitude expression applies on the side of the threshold for which the quantity under the square root is positive.

\section{Numerical Validation \label{sec:simulations}}

To assess the accuracy of the weakly nonlinear theory, we compare the analytical predictions (\ref{eq:tipamplitude}) and (\ref{eq:tipfrequency}) with direct numerical simulations of the full nonlinear Cosserat rod equations. Numerical simulations were performed for the four representative parameter sets listed in Table \ref{tab:parameters}, using the methodology outlined in \citep{Warda2025}. Details of the numerical implementation are provided in Appendix \ref{sec:numerics}. The parameter sets span different stretch and shear stiffness ratios and include both the lower and upper stability thresholds of the reentrant instability window. The saturated oscillation amplitude and frequency were extracted from the
long-time periodic motion of the rod tip after transient decay. The tip
amplitude $A_{\mathrm{tip}}$ is defined as one half of the peak-to-peak
transverse displacement of the rod tip over one oscillation period,
while the oscillation frequency was determined from the dominant peak of the Fourier spectrum of the long-time tip displacement. Numerical simulations were continued until the oscillation amplitude and
period became stationary. The integration time $\mathcal{T}=5$ was found to be sufficient for all parameter sets considered, including those closest to the instability threshold.

 \onecolumngrid
\begin{widetext}
    \begin{figure}[H]
\centering
\includegraphics[scale=0.45]{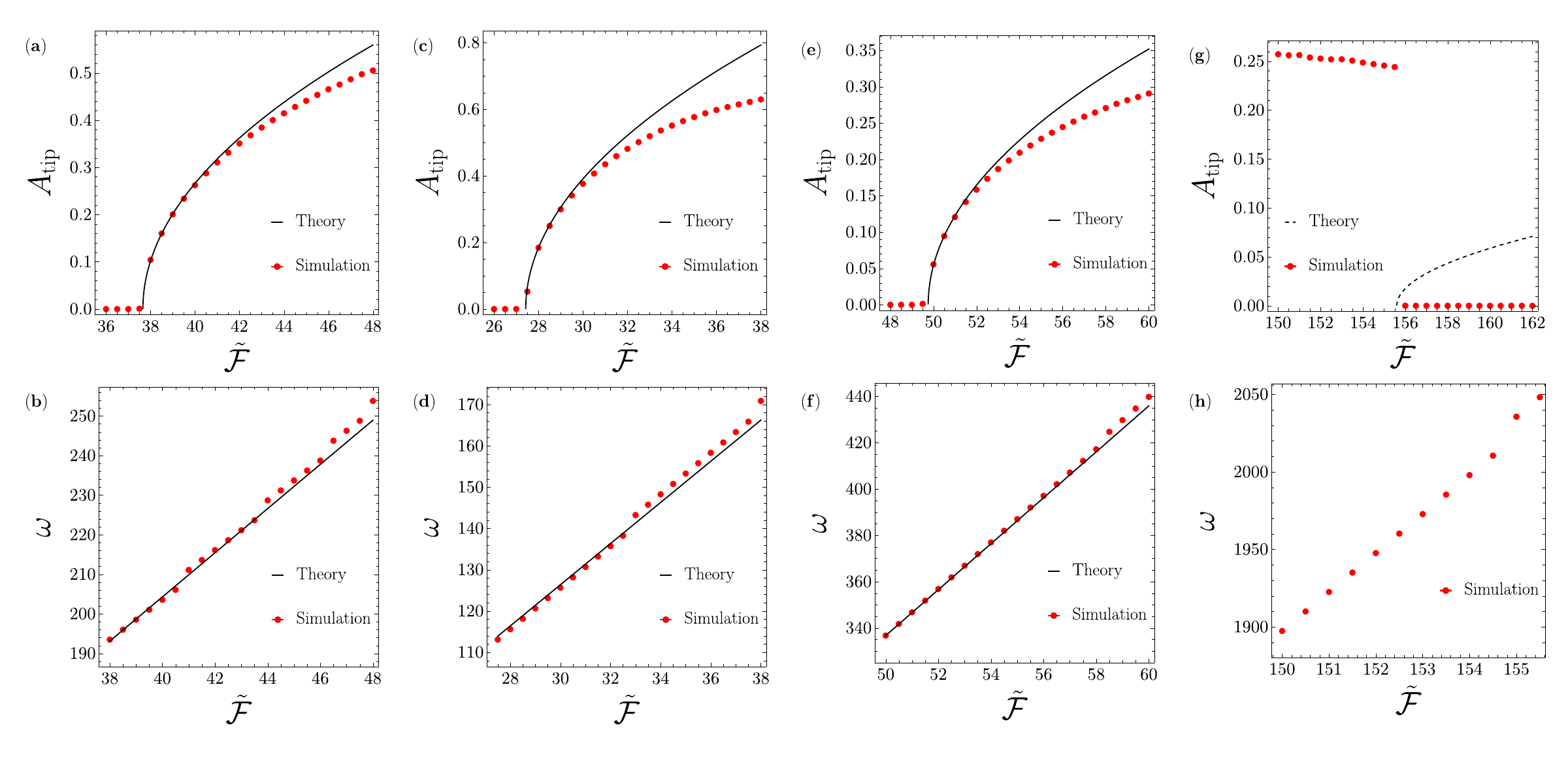}
\caption{\label{fig:comparison}
Comparison between the weakly nonlinear theory and direct nonlinear
simulations for four representative constitutive parameter sets listed
in Table \ref{tab:parameters}. Here simulation data for $A_{\mathrm{tip}}$ denotes one half of the peak-to-peak transverse displacement of the rod tip measured after convergence to the periodic state, while $\omega$ data denotes the dominant peak of the Fourier spectrum of the long-time tip displacement. Each column corresponds to one parameter set, with the top row ((a), (c), (e), and (g)) showing the saturated tip oscillation amplitude $A_{\mathrm{tip}}$ as a function of the follower force $\tilde{\mathcal F}$, and the bottom row ((b), (d), (f), and (h))
showing the corresponding oscillation frequency $\omega$. Black curves denote the weakly nonlinear predictions given by (\ref{eq:tipamplitude}) and (\ref{eq:tipfrequency}), while red symbols are obtained from direct
numerical simulations of the full nonlinear Cosserat-rod equations.
Panels (a)--(f) correspond to supercritical Hopf bifurcations, for
which the theory accurately predicts the square-root growth of the
saturated oscillation amplitude together with the linear frequency
correction near threshold. Panels (g) and (h) correspond to the upper
stability threshold of the reentrant instability window, where the
Landau coefficients predict a subcritical Hopf bifurcation. The dashed
curve in (g) denotes the unstable branch predicted by the cubic
Stuart-Landau equation.
}
\end{figure}

\begin{table}[H]
  \centering
  \begin{ruledtabular}
    \begin{tabular}{cccccccc}
      Figure & $\tilde{k}_{1}$ & $\tilde{k}_{2}$ & $\tilde{\gamma}_{3}$ & $\tilde{\mathcal{F}}_c$& $\omega_c$& $\alpha$ & $\beta$ \\
      \colrule
      (a, b) & $10^{4}$        & $10^{4}$        & $10^{-4}$ & $37.66$ & $191.21$  & $-971.79 + 29.09i$ &  $7.35 + 5.36 i$\\
      (c, d) & $10^{4}$        & $10^{2}$        & $10^{-2}$ & $27.43$ & $113.55$& $-466.06+57.49 i$& 6.93+4.13 i  \\
      (e, f) & $2\times 10^{2}$& $10^{4}$        & $10^{-4}$ & $49.76$ & 334.32& $-2183.57 + 252.12i$ & $6.60 + 9.16 i$ \\
      (g, h) & $2\times 10^{2}$        & $10^{4}$        & $10^{-4}$ & $155.60$ & $3367.28$ & $328902 - 307479 i$ & $-64.59 + 88.26i$ \\
    \end{tabular}
  \end{ruledtabular}
  \caption{ Material parameters, critical follower forces, Hopf frequencies, and corresponding Landau coefficients used to generate the weakly nonlinear predictions shown in Fig. \ref{fig:comparison}. The Landau coefficients $\alpha$ and $\beta$ are obtained from (\ref{eq:alphabetaformulas}) and determine both the criticality of the Hopf bifurcation and the analytical predictions for the saturated oscillation amplitude and nonlinear frequency correction. Numerical values are reported to two decimal places.}\label{tab:parameters}
\end{table}

\end{widetext}
\twocolumngrid

Figure \ref{fig:comparison} compares the resulting simulation data with the weakly nonlinear predictions. Panels (a)--(f) correspond to lower stability thresholds for which $\operatorname{Re}(\alpha)<0$. The Hopf bifurcation is therefore supercritical, and a stable finite-amplitude branch exists on the linearly unstable side of the threshold. In each case, the predicted tip amplitude is given by (\ref{eq:tipamplitude}) with the square root evaluated on the side for which its argument is
positive. The simulations reproduce the predicted square-root growth
close to onset. This agreement confirms that the cubic nonlinearity
retained in the Stuart-Landau equation captures the leading saturation
mechanism of the follower force instability.

The agreement is strongest sufficiently close to the critical force.
Further from threshold, the numerical amplitudes increasingly depart from
the weakly nonlinear curves, with the analytical prediction overestimating the saturated amplitude. Such deviations are expected
because the asymptotic expansion assumes
$|\tilde{\mathcal F}-\tilde{\mathcal F}_{c}|\ll 1$. At larger distances
from threshold, higher-order nonlinearities, higher temporal harmonics,
and corrections to the leading spatial eigenmode become increasingly
important.

The lower panels of the first three columns compare the corresponding
oscillation frequencies. The weakly nonlinear theory predicts the linear
correction given by (\ref{eq:tipfrequency}). For all three supercritical cases, the simulations exhibit an approximately linear variation of the frequency near threshold, with slopes in good agreement with the analytical prediction. The frequency is typically predicted over a wider interval than the amplitude, although small systematic deviations again become visible as the system moves further from the asymptotic regime.

Panels (g) and (h) examine the upper boundary of the reentrant instability
window for $\tilde{k}_{1}=2\times10^{2}$, $\tilde{k}_{2}=10^{4}$, and $\tilde{\gamma}_{3}=10^{-4}$. At this threshold, $\operatorname{Re}(\alpha)>0$, whereas $\operatorname{Re}(\beta)<0$. Since increasing
$\tilde{\mathcal F}$ through the upper threshold stabilizes the straight
state, the linearly unstable side corresponds to $\tilde{\mathcal F}<\tilde{\mathcal F}_{c}$. The nonzero branch predicted by the cubic normal form instead lies on the linearly stable side, $\tilde{\mathcal F}>\tilde{\mathcal F}_{c}$, and is radially unstable. The upper transition is therefore subcritical.

The dashed curve in Fig. \ref{fig:comparison}(g) represents this unstable small-amplitude branch. It should not be interpreted as a prediction for the stable oscillations obtained by direct time integration, since generic
simulations cannot converge to a radially unstable periodic orbit. The
finite-amplitude oscillations observed below the upper threshold do not decrease continuously to zero as the threshold is approached. Their eventual fate is controlled
by nonlinear terms beyond cubic order and cannot be determined from the
Stuart-Landau equation derived here. In particular, the cubic theory does
not predict the location of a possible fold of periodic orbits or the
extent of any associated hysteretic interval.

For this reason, no stable weakly nonlinear frequency branch is plotted in
Fig. \ref{fig:comparison}(h). The simulation data nevertheless show the frequency of the stable finite-amplitude oscillations on the unstable side of the upper threshold. Because these oscillations do not approach zero amplitude as $\tilde{\mathcal F}\to\tilde{\mathcal F}_{c}^{-}$, they lie outside the local small-amplitude regime described by the cubic normal form.

Taken together, the four cases demonstrate that the Landau coefficients
provide more than an amplitude fit. Their real parts determine whether
the transition is supercritical or subcritical, while their imaginary
parts predict the leading nonlinear frequency shift. The reduced theory
accurately describes the stable oscillatory branch near the three
supercritical lower thresholds and correctly identifies the qualitatively
different subcritical structure of the upper restabilization threshold. Finally, it should be noted that the Landau coefficient $\beta$ admits a direct consistency check against the the variation of
the critical eigenvalue obtained from the linear stability problem. This is presented in Appendix \ref{sec:consistency}.

\section{Discussion and Conclusion}

In this work, we developed a weakly nonlinear theory for the onset and
saturation of follower-force-driven oscillations in a planar Cosserat rod
immersed in a viscous fluid. Building on the linear stability analysis of
our preceding work, we considered follower forces near a Hopf bifurcation
of the compressed straight state and performed a multiple-scale expansion
about the corresponding critical equilibrium. The resulting asymptotic
hierarchy naturally separates the leading oscillatory dynamics from the
slow nonlinear modulation responsible for amplitude saturation.

At leading order, the dynamics is governed by the critical eigenmode of a
non-self-adjoint linear operator arising from the follower force boundary
condition. At second order, quadratic nonlinearities generate static and
second-harmonic longitudinal correction fields associated with stretch and
shear, which feed back into the transverse dynamics at cubic order.
Projection of the resonant forcing onto the adjoint critical eigenmode
then yields a Stuart-Landau amplitude equation governing the slow
evolution of the Hopf mode. The resulting Landau coefficients are obtained
explicitly in terms of the critical eigenmode, its adjoint, and the
quadratic correction fields.

The reduced amplitude equation provides an analytical normal-form
description of the follower force instability in geometrically exact
Cosserat rods. Its real coefficients determine whether the Hopf
bifurcation is supercritical or subcritical, while its imaginary
coefficients determine the leading nonlinear frequency correction. For
supercritical bifurcations, the theory predicts the square-root growth of
the saturated oscillation amplitude together with a frequency correction
linear in the distance from threshold. Comparison with direct numerical
simulations demonstrates excellent agreement near onset across several
representative material parameter regimes. Furthermore, application of the
same framework to the upper boundary of the finite instability window
correctly identifies the restabilization transition as subcritical,
highlighting the ability of the weakly nonlinear theory to distinguish the
qualitatively different nonlinear structures associated with the two
stability boundaries.

It is instructive to place these results in the context of recent weakly
nonlinear analyses of follower force instabilities. Schnitzer \citep{Schnitzer2025} considered an inextensible, unshearable Kirchhoff filament undergoing a symmetry-induced double Hopf bifurcation in three dimensions, leading to a
higher-dimensional amplitude equation describing the emergence of whirling
states. By contrast, the present work considers geometrically exact planar
Cosserat rods with extensibility and shear, for which the Hopf
bifurcation is non-degenerate and the leading-order dynamics is described
by a Stuart-Landau equation. The additional Cosserat degrees of freedom
introduce longitudinal correction fields and quadratic couplings absent
from the Kirchhoff formulation, while the finite instability window unique
to the Cosserat model enables weakly nonlinear analysis at both the onset
and restabilization boundaries.

The resulting amplitude equation provides a reduced normal form description of the flutter instability in the planar Cosserat-rod setting. For the lower stability thresholds considered here, the resulting amplitude equation predicts supercritical Hopf bifurcations, with a steady-state oscillation amplitude proportional to the square root of the distance from threshold and a frequency correction linear in that distance. At the upper restabilization threshold, the same framework instead predicts a subcritical Hopf bifurcation and an unstable small-amplitude periodic branch. Beyond its asymptotic value, the reduced model provides a compact description suitable for parameter sweeps and control-oriented modeling of pressure-driven soft robotic arms in viscous environments.

Several extensions naturally suggest themselves. The weakly nonlinear
framework developed here may be applied to more general active Cosserat-rod
models, including recently proposed models of nonreciprocal active slender
bodies in which strain-dependent force and torque densities generate fixed
points, limit cycles, and chaotic attractors in shape space \citep{Nemeth2016}. Such an
analysis would provide analytical normal-form descriptions of the pitchfork,
Hopf, and other bifurcations observed numerically, and clarify the role of
nonlinear mode interactions in the emergence of autonomous locomotion and
more complex dynamical states. Extensions to fully three-dimensional
Cosserat rods with twist could reveal symmetry-induced degenerate Hopf
bifurcations and higher-dimensional amplitude equations, while
incorporating nonlocal hydrodynamic interactions beyond local Stokes drag
would provide a more realistic description of slender bodies in viscous
fluids. More broadly, the analytical framework developed here provides a
foundation for studying nonlinear nonconservative instabilities in
geometrically exact rods and for developing reduced-order descriptions of
active and soft robotic slender structures operating at low Reynolds
number.

\begin{acknowledgments}
I thank Professor Ronojoy Adhikari for suggesting the problem. I thank Professors M. C. Payne, M. E. Cates, Ronojoy Adhikari, and Dr. A. Souslov for facilitating a collaboration between the Cavendish Laboratory and the Department of Applied Mathematics and Theoretical
Physics. I acknowledge a PhD studentship from the EPSRC (award number W108115D) and UKRI.
\end{acknowledgments}

\section*{Data Availability}

The numerical simulation data and Mathematica notebooks used to generate the results of this study are available from the corresponding author upon reasonable request.

\appendix
\clearpage
\onecolumngrid

\section{Multiple-scale Analysis \label{sec:multiplescaleanalysis}}
We proceed with the method of multiple scales \citep{Hinch1991} as follows. We consider values of the follower force near the stability threshold
\begin{equation}
    \tilde{\mathcal{F}} = \tilde{\mathcal{F}}_c + \epsilon^2 \Delta.
\end{equation}
We introduce a fast time scale $\tau$ and a slow time scale $T$ related on the physical diagonal by
\begin{equation}
    \tau(t) =t,\quad T(t) = \epsilon^2 t.
    \label{eq:timescales}
\end{equation}
The fast time $\tau$ resolves oscillations at the Hopf frequency $\omega_c$, while the slow time $T$ captures the modulation of the oscillation amplitude near threshold. In what follows, $\tau$ and $T$ are treated as independent variables, and the dependence on $T$ is determined by eliminating secular growth in the perturbation expansion. Accordingly, we promote all quantities in our field theory to be functions of $u$ and the two independent temporal coordinates $\tau$ and $T$. For any quantity $\mathcal{A}(u,t)$ in our field theory, we write it as $\mathcal{A}(u, \tau, T)$. We carry out a perturbative expansion in $\epsilon$ as
\begin{equation}
    \mathcal{A}(u, \tau, T) = \sum_{n=0}^\infty \mathcal{A}^{(n)}(u, \tau, T) \,  \epsilon^n,
    \label{eq:perturbationtheory}
\end{equation}
where $\mathcal{A}^{(n)}$ is the $\mathcal{O}(\epsilon^n)$ correction to $\mathcal{A}$ about the base state (\ref{eq:basestate}) corresponding to $\tilde{\mathcal{F}}= \tilde{\mathcal{F}}_c$. We expand about the base state at the critical follower force $\tilde{\mathcal{F}} = \tilde{\mathcal{F}}_c$. In particular, the zeroth-order fields correspond to the straight compressed configuration with uniform internal force and vanishing moment as given in (\ref{eq:basestate}). The higher-order corrections describe small transverse deflections, longitudinal adjustments, and curvature variations induced by the follower force instability.

The definitions in (\ref{eq:timescales}) imply that the time derivative operator $\partial_t$ transforms as
\begin{equation}
    \partial_t \to \partial_\tau + \epsilon^2 \partial_T.
    \label{eq:timederivative}
\end{equation}
The slow-time derivative first appears at $\mathcal{O}(\epsilon^3)$ in the kinematics, consistent with the fact that near threshold the amplitude of the critical mode evolves on the long time scale $t = \mathcal{O}(\epsilon^{-2})$. As a result, it is sufficient to carry the perturbation expansion through cubic order in $\epsilon$ in order to derive the amplitude equation governing the Hopf bifurcation. We collect terms up to cubic order in $\epsilon$, which will be sufficient to determine the leading-order amplitude dynamics near the Hopf bifurcation. Substituting the expansions (\ref{eq:perturbationtheory}) into the exact kinematic definitions (\ref{eq:definitions}) and expanding in powers of $\epsilon$, we obtain the corrections to the strain variables $(h_1, h_2, \Pi)$ and velocity variables $(v_1, v_2, \Omega)$ expressed in terms of the corrections to $(x, y, \theta)$. For clarity we list these expansions explicitly through $\mathcal{O}(\epsilon^3)$. 

At $\mathcal{O}(\epsilon)$, we find
\begin{equation}
        h_1^{(1)} = \partial_u x^{(1)},\quad
        h_2^{(1)} = \partial_u y^{(1)} - \nu \theta^{(1)}, \quad
        \Pi^{(1)} = \partial_u \theta^{(1)},
    \label{eq:strainlinear}
\end{equation}
and
\begin{equation}
        v_1^{(1)} = \partial_\tau x^{(1)},\quad
        v_2^{(1)} = \partial_\tau y^{(1)},\quad
        \Omega^{(1)} = \partial_\tau \theta^{(1)}.
    \label{eq:velocitylinear}
\end{equation}
At this order the kinematics is linear, and the resulting governing equations recover the linear stability problem analysed in Sec. \ref{sec:hopfbifurcation}. At $\mathcal{O}(\epsilon^2)$, we find
\begin{equation}
        h_1^{(2)} = \partial_u x^{(2)} + \theta^{(1)} h_2^{(1)} + \frac{1}{2} \nu (\theta^{(1)})^2,\quad
        h_2^{(2)} = \partial_u y^{(2)} - \nu \theta^{(2)} - \theta^{(1)} \partial_u x^{(1)},\quad
        \Pi^{(2)} = \partial_u \theta^{(2)}
    \label{eq:strainquadratic}
\end{equation}
and
\begin{equation}
        v_1^{(2)} = \partial_\tau x^{(2)} + \theta^{(1)} \partial_\tau y^{(1)},\quad
        v_2^{(2)} = \partial_\tau y^{(2)} - \theta^{(1)} \partial_\tau x^{(1)},\quad
        \Omega^{(2)} = \partial_\tau \theta^{(2)}.
    \label{eq:velocityquadratic}
\end{equation}
At this quadratic order, the kinematic relations already contain quadratic products of the leading-order fields. When substituted into the dynamics, these terms act as inhomogeneous forcing for the $\mathcal{O}(\epsilon^2)$ correction fields. Finally, at $\mathcal{O}(\epsilon^3)$, we find
\begin{equation}
    \begin{aligned}
        h_1^{(3)} &= \partial_u x^{(3)} - \nu \theta^{(1)} \theta^{(2)}  - \frac{1}{2} (\theta^{(1)})^2 \partial_u x^{(1)} + \theta^{(1)} \partial_u y^{(2)} + \theta^{(2)} \partial_u y^{(1)}\\
        h_2^{(3)} &= \partial_u y^{(3)} - \nu \left(\theta^{(3)} - \frac{1}{6} (\theta^{(1)})^3 \right) - \frac{1}{2} (\theta^{(1)})^2 \partial_u y^{(1)} -\theta^{(1)} \partial_u x^{(2)} - \theta^{(2)} \partial_u x^{(1)}\\
        \Pi^{(3)} &= \partial_u \theta^{(3)}
    \end{aligned}
    \label{eq:straincubic}
\end{equation}
and
\begin{equation}
    \begin{aligned}
        v_1^{(3)} &= \partial_\tau x^{(3)} + \partial_T x^{(1)} -\frac{1}{2} (\theta^{(1)})^2 \partial_\tau x^{(1)} +  \theta^{(1)} \partial_\tau y^{(2)} + \theta^{(2)} \partial_\tau y^{(1)}\\
        v_2^{(3)} &= \partial_\tau y^{(3)} + \partial_T y^{(1)} - \frac{1}{2} (\theta^{(1)})^2 \partial_\tau y^{(1)} - \theta^{(1)} \partial_\tau x^{(2)}  - \theta^{(2)} \partial_\tau x^{(1)}\\
        \Omega^{(3)} &= \partial_\tau \theta^{(3)} + \partial_T \theta^{(1)}
    \end{aligned}
    \label{eq:velocitycubic}
\end{equation}
At this cubic order, the slow-time derivatives $\partial_T y^{(1)}$ and $\partial_T \theta^{(1)}$ enter the kinematics. This is the first order at which the modulation of the leading-order oscillation appears, and it is therefore the order at which a solvability condition yields the amplitude equation.

We now develop the weakly nonlinear theory by substituting the above kinematic expansions into the governing dynamical equations
\begin{equation}
\gamma_1 v_1 = \partial_uF_1 - \Pi F_2,\quad\gamma_2 v_2 = \partial_uF_2 + \Pi F_1,\quad \gamma_3 \Omega = \partial_uM + h_1F_2 - h_2F_1,
    \label{eq:dynamicsfull}
\end{equation}
together with the boundary conditions (\ref{eq:clamping}) and (\ref{eq:follower}). The stress corrections $(F_1^{(n)}, F_2^{(n)}, M^{(n)})$ are trivially related to the strain corrections $(h_1^{(n)}, h_2^{(n)}, \Pi^{(n)})$ through the constitutive law (\ref{eq:constitutive}). Collecting terms at successive orders in $\epsilon$ yields a hierarchy of forced linear boundary-value problems. The $\mathcal{O}(\epsilon)$ problem reproduces the Hopf eigenmode at the threshold, the $\mathcal{O}(\epsilon^2)$ problem determines the quadratic corrections, and the $\mathcal{O}(\epsilon^3)$ problem provides a solvability condition that closes as a Stuart-Landau equation for the slow evolution of the mode amplitude.

\section{Weakly Nonlinear Theory and Landau Coefficients \label{sec:weaklynonlinear}}
\subsection{$\mathcal{O}(\epsilon)$ Theory}

At leading order, we expand the full overdamped Cosserat-rod dynamics
(\ref{eq:dynamicsfull}) about the compressed straight base state
(\ref{eq:basestate}) at the critical follower force
$\tilde{\mathcal{F}}=\tilde{\mathcal{F}}_c$. Since the detuning from
threshold enters at $O(\epsilon^2)$ via
$\tilde{\mathcal{F}}=\tilde{\mathcal{F}}_c+\epsilon^2\Delta$, the
$O(\epsilon)$ problem is evaluated at the critical value
$\mathcal{F}=\mathcal{F}_c$.

The $\mathcal{O}(\epsilon)$ expansion of (\ref{eq:dynamicsfull}) together
with the boundary conditions (\ref{eq:clamping}) and (\ref{eq:follower})
yields the linear system
\begin{equation}
        \tilde{\gamma}_1 v_1^{(1)} = \partial_u F_1^{(1)},\quad
        v_2^{(1)} = \partial_u F_2^{(1)} - \tilde{\mathcal{F}}_c \Pi^{(1)},\quad
        \tilde{\gamma}_3 \Omega^{(1)} = \partial_u M^{(1)} + \nu F_2^{(1)} + \tilde{\mathcal{F}}_c h_2^{(1)}.
    \label{eq:dynamicslinear}
\end{equation}
The clamped boundary conditions at $u=0$ imply
\begin{equation}
    x^{(1)} = y^{(1)} = \theta^{(1)} = 0 \quad \mathrm{at} \,\, u=0,
\end{equation}
while the follower force boundary conditions at $u=1$ linearize to the
homogeneous traction conditions
\begin{equation}
    F_1^{(1)} = F_2^{(1)} = M^{(1)} = 0 \quad \mathrm{at} \,\, u=1.
    \label{eq:followerlinear}
\end{equation}
Substituting the leading-order kinematic relations
(\ref{eq:strainlinear}) and (\ref{eq:velocitylinear}) and using the
linear constitutive law (\ref{eq:constitutive}), we obtain the closed
system for the perturbation fields $(x^{(1)},y^{(1)},\theta^{(1)})$,
\begin{equation}
            \tilde{\gamma}_1 \partial_\tau x^{(1)} = \tilde{k}_1 \partial_u^2 x^{(1)}, \quad
           \partial_\tau y^{(1)} =  \tilde{k}_2\partial_u^2 y^{(1)} - \mu \partial_u \theta^{(1)},\quad
           \tilde{\gamma}_3 \partial_\tau \theta^{(1)} = \partial_u^2 \theta^{(1)} + \mu \bigl( \partial_u y^{(1)} - \nu \theta^{(1)}\bigr),
\end{equation}
with clamping conditions
\begin{equation}
    x^{(1)} = y^{(1)} = \theta^{(1)} = 0 \quad \mathrm{at} \,\, u=0,
\end{equation}
and free-end conditions
\begin{equation}
    \partial_ux^{(1)} = \partial_uy^{(1)} - \nu \theta^{(1)} = \partial_u\theta^{(1)} = 0
    \quad \mathrm{at} \,\, u=1.
\end{equation}
This is precisely the linear stability problem summarized in
(\ref{eq:linearization}), evaluated at the Hopf bifurcation threshold.

At $\tilde{\mathcal{F}}=\tilde{\mathcal{F}}_c$, the spectrum of the
linear operator $\mathcal{L}$ contains a critical eigenvalue
$i\omega_c$ associated with a transverse eigenmode
$(0,Y(u),\Theta(u))$. The general $O(\epsilon)$ solution therefore takes
the form of a neutrally stable oscillation at the Hopf frequency,
with a complex amplitude that is allowed to vary on the slow time scale
$T=\epsilon^2 t$,
\begin{equation}
        x^{(1)}(u, \tau, T) = 0,\quad
        y^{(1)}(u, \tau, T) = A(T) Y(u) e^{i\omega_c\tau} + \mathrm{c.c.},\quad
        \theta^{(1)}(u, \tau, T) = A(T) \Theta(u) e^{i\omega_c\tau} + \mathrm{c.c.}
        \label{eq:solutionlinear}
\end{equation}
We normalize the eigenmode such that
$Y(1)=1$, so that the complex amplitude $A(T)$ coincides with the
leading-order tip displacement in the transverse direction. In what follows, we introduce the shear shape function $H \equiv Y' - \nu \Theta$ associated with the critical eigenmode for convenience. 

\subsection{$\mathcal{O}(\epsilon^2)$ Theory}

At $O(\epsilon^2)$, the governing equations contain quadratic products of
the leading-order oscillatory fields. Since the $O(\epsilon)$ solution
(\ref{eq:solutionlinear}) is purely transverse and oscillatory at the Hopf
frequency $\omega_c$, the quadratic terms generate forcing at the mean
frequency $0$ and at the second harmonic $2\omega_c$. Importantly, no
terms proportional to $e^{i\omega_c\tau}$ appear at this order, and thus
the $O(\epsilon^2)$ problem does not require a solvability condition.
Instead, the role of the $O(\epsilon^2)$ theory is to compute the
non-resonant correction fields that will enter the cubic-order forcing in
the $O(\epsilon^3)$ problem.

The $\mathcal{O}(\epsilon^2)$ expansion of (\ref{eq:dynamicsfull}) and the
boundary conditions (\ref{eq:clamping}) and (\ref{eq:follower}), together
with the transverse structure of the solution (\ref{eq:solutionlinear}),
yields
\begin{equation}
        \tilde{\gamma}_1 v_1^{(2)} = \partial_u F_1^{(2)} - \Pi^{(1)} F_2^{(1)},\quad
       v_2^{(2)} = \partial_u F_2^{(2)} - \tilde{\mathcal{F}}_c \Pi^{(2)}, \quad
        \tilde{\gamma}_3 \Omega^{(2)} = \partial_u M^{(2)} + \nu F_2^{(2)} + \tilde{\mathcal{F}}_c h_2^{(2)}.
    \label{eq:dynamicsquadratic}
\end{equation}
The clamping conditions remain homogeneous,
\begin{equation}
    x^{(2)} = y^{(2)} = \theta^{(2)} = 0 \quad \mathrm{at} \,\, u=0,
\end{equation}
while the follower force boundary condition introduces the detuning
parameter $\Delta$ through the expansion
$\tilde{\mathcal{F}}=\tilde{\mathcal{F}}_c+\epsilon^2\Delta$, giving
\begin{equation}
    F_1^{(2)} = -\Delta, \quad  F_2^{(2)}= M^{(2)} = 0 \quad \mathrm{at} \,\, u=1.
    \label{eq:followerquadratic}
\end{equation}
Substituting the $O(\epsilon^2)$ kinematics (\ref{eq:strainquadratic}) and
(\ref{eq:velocityquadratic}) into (\ref{eq:dynamicsquadratic}) and using
the constitutive law (\ref{eq:constitutive}), we obtain
\begin{equation}
            \tilde{\gamma}_1 \partial_\tau x^{(2)} = \tilde{k}_1 \partial_u^2 x^{(2)} + f_1^{(2)},\quad
             \partial_\tau y^{(2)} =  \tilde{k}_2\partial_u^2 y^{(2)} - \mu \partial_u \theta^{(2)},\quad
            \tilde{\gamma}_3 \partial_\tau \theta^{(2)} = \partial_u^2 \theta^{(2)} + \mu \bigl( \partial_u y^{(2)} - \nu \theta^{(2)}\bigr),
       \label{eq:quadratictheory}
\end{equation}
with homogeneous clamping
\begin{equation}
    x^{(2)} = y^{(2)} = \theta^{(2)} = 0 \quad \mathrm{at} \,\, u=0,
    \label{eq:quadraticclamping}
\end{equation}
and free-end conditions
\begin{equation}
    \partial_ux^{(2)} = b_1^{(2)}, \quad \partial_uy^{(2)} - \nu \theta^{(2)} = \partial_u\theta^{(2)} = 0
    \quad \mathrm{at} \,\, u=1.
    \label{eq:quadratictheoryBC}
\end{equation}
Here $f_1^{(2)}$ and $b_1^{(2)}$ are inhomogeneous forcing terms that
depend on the leading-order oscillatory solution
(\ref{eq:solutionlinear}). In particular, evaluating the quadratic terms
explicitly yields
\begin{equation}
    \begin{aligned}
        f_1^{(2)} &= \theta^{(1)} \left(\tilde{k}_1 \partial_u^2 y^{(1)} - \tilde{\gamma}_1 \partial_\tau y^{(1)} \right) + (\tilde{k}_1 - \tilde{k}_2) h_2^{(1)} \partial_u \theta^{(1)}\\
        &= A^2 \left[\Theta(\tilde{k}_1 Y'' - i \omega_c \tilde{\gamma}_1 Y) + (\tilde{k}_1 - \tilde{k}_2) H \Theta' \right] e^{2 i\omega_c \tau}+ |A|^2\left[ \Theta^*(\tilde{k}_1 Y'' - i \omega_c \tilde{\gamma}_1 Y) + (\tilde{k}_1 - \tilde{k}_2) H^* \Theta'\right]
        + \mathrm{c.c.},
    \end{aligned}
    \label{eq:longitudinalforcing}
\end{equation}
and
\begin{equation}
    \begin{aligned}
        b_1^{(2)} &= -\frac{1}{2} \nu (\theta^{(1)}(u=1))^2 - \frac{\Delta}{\tilde{k}_1}\\
        &= -\frac{1}{2} A^2 \nu \Theta(1)^2 e^{2i\omega_c\tau}
        - \frac{1}{2}|A|^2 \nu |\Theta(1)|^2
        - \frac{\Delta}{2\tilde{k}_1}
        + \mathrm{c.c.}.
    \end{aligned}
    \label{eq:longitudinalforcingBC}
\end{equation}
We therefore see that the $O(\epsilon^2)$ problem is driven only through
the longitudinal equation for $x^{(2)}$ and its boundary condition at
$u=1$, while the transverse fields $y^{(2)}$ and $\theta^{(2)}$ satisfy
the homogeneous linear problem. Since the forcing contains only the
frequencies $0$ and $2\omega_c$, the transverse homogeneous problem admits
the trivial solution $y^{(2)}=\theta^{(2)}=0$ without loss of generality.
We adopt this choice in what follows, so that the only nontrivial
$O(\epsilon^2)$ correction is a longitudinal displacement $x^{(2)}$.

Motivated by the frequency content of (\ref{eq:longitudinalforcing}) and
(\ref{eq:longitudinalforcingBC}), we seek a particular solution in the
form
\begin{equation}
    x^{(2)} = A^2 X_2(u) e^{2i\omega_c\tau} + |A|^2 X_0(u) - \frac{\Delta}{2\tilde{k}_1}u + \mathrm{c.c.}
\end{equation}
Substituting this ansatz into the longitudinal equation in
(\ref{eq:quadratictheory}) yields a pair of inhomogeneous ordinary
differential equations for the mode shapes $X_0$ and $X_2$,
\begin{equation}
        \tilde{k}_1 X_0'' = \Theta^* (i\omega_c \tilde{\gamma}_1 Y - \tilde{k}_1 Y'') +  (\tilde{k}_2 - \tilde{k}_1) H^* \Theta',\quad
        \tilde{k}_1 X_2'' = 2 i \omega_c \tilde{\gamma}_1 X_2 + \Theta(i\omega_c \tilde{\gamma}_1 Y - \tilde{k}_1 Y'')+ (\tilde{k}_2 - \tilde{k}_1) H\Theta',
    \label{eq:longitudinalODE}
\end{equation}
subject to the mixed boundary conditions
\begin{equation}
        X_0(0) = 0,\quad
        X_0'(1) = - \frac{1}{2} \nu |\Theta(1)|^2,\quad
        X_2(0) = 0,\quad
        X_2'(1) = -\frac{1}{2} \nu \Theta(1)^2.
    \label{eq:ODEBCs}
\end{equation}
The system of boundary value problems (\ref{eq:longitudinalODE}) and (\ref{eq:ODEBCs}) may also be solved analytically or numerically.

For later use in the cubic-order forcing terms, it is convenient to
express the longitudinal stress correction $F_1^{(2)}$ in a form
consistent with the frequency decomposition of $x^{(2)}$,
\begin{equation}
    F_1^{(2)} = A^2 P_2 e^{2i\omega_c\tau} + |A|^2 P_0 -\frac{\Delta}{2} + \mathrm{c.c.},
    \label{eq:pdefinitions}
\end{equation}
where $P_0$ and $P_2$ are related to $X_0$ and $X_2$ through the
constitutive law as
\begin{equation}
        P_0 = \tilde{k}_1\left(X_0' + \Theta^*H + \frac{1}{2} \nu |\Theta|^2 \right),\quad
        P_2 = \tilde{k}_1 \left(X_2' + \Theta H + \frac{1}{2} \nu \Theta^2 \right).
\end{equation}
Collecting the results of this subsection, the $O(\epsilon^2)$ correction
fields may be written as
\begin{equation}
    x^{(2)} = A^2 X_2 e^{2i\omega_c\tau} + |A|^2 X_0 - \frac{\Delta}{2\tilde{k}_1}u + \mathrm{c.c.},\quad
    y^{(2)} = 0,\quad
    \theta^{(2)} = 0.
\end{equation}

\subsection{$\mathcal{O}(\epsilon^3)$ Theory}

At $O(\epsilon^3)$, the multiple-scale structure first enters explicitly
through the slow-time derivatives $\partial_T y^{(1)}$ and
$\partial_T\theta^{(1)}$ in the kinematics (\ref{eq:velocitycubic}).
Consequently, this order contains the terms proportional to $\mathrm{d}A/\mathrm{d}T$
that ultimately determine the amplitude evolution. In addition, products
of the $O(\epsilon)$ and $O(\epsilon^2)$ fields generate cubic forcing at
the Hopf frequency $e^{i\omega_c\tau}$. As in standard Hopf normal-form
derivations, the resonant part of this forcing must satisfy a solvability
condition, yielding the Stuart-Landau equation.

The $\mathcal{O}(\epsilon^3)$ expansion of the dynamics
(\ref{eq:dynamicsfull}) and the boundary conditions
(\ref{eq:clamping})--(\ref{eq:follower}) yields
\begin{equation}
\begin{aligned}
\tilde{\gamma}_1 v_1^{(3)} &= \partial_u F_1^{(3)},\\
v_2^{(3)} &= \partial_u F_2^{(3)} - \tilde{\mathcal{F}}_c \Pi^{(3)} + \Pi^{(1)} F_1^{(2)},\\
\tilde{\gamma}_3 \Omega^{(3)} &= \partial_u M^{(3)} + \nu F_2^{(3)} + \tilde{\mathcal{F}}_c h_2^{(3)}
- h_2^{(1)} F_1^{(2)} + h_1^{(2)} F_2^{(1)}.
\end{aligned}
\label{eq:dynamicscubic}
\end{equation}
The clamping conditions remain homogeneous,
\begin{equation}
x^{(3)} = y^{(3)} = \theta^{(3)} = 0 \quad \mathrm{at}\quad u=0,
\label{eq:clampingcubic}
\end{equation}
and the $u=1$ boundary conditions obtained from the follower force
expansion take the form
\begin{equation}
\begin{aligned}
F_1^{(3)} = F_2^{(3)} = M^{(3)} &= 0 \quad \mathrm{at}\quad u=1,
\end{aligned}
\label{eq:followercubic_stress}
\end{equation}
which may be written in terms of the $cubic$-order corrections at $u = 1$ as
\begin{equation}
\partial_u x^{(3)} = 0,\quad
\partial_u y^{(3)} - \nu\!\left(\theta^{(3)}-\frac{1}{6}(\theta^{(1)})^3\right) =
-\frac{\Delta}{\tilde{k}_1}\,\theta^{(1)}(u=1),\quad
\partial_u\!\left(\theta^{(3)}-\frac{1}{6}(\theta^{(1)})^3\right) = 0.
\label{eq:followercubic_geom}
\end{equation}

The inhomogeneous term in (\ref{eq:followercubic_geom}) is proportional to
$\Delta A(T)e^{i\omega_c\tau}$ and therefore has the same temporal
dependence as the critical mode. If left in the boundary conditions, this
term would appear as a resonant boundary forcing in the $O(\epsilon^3)$
problem and would obscure the standard solvability structure. To
homogenize the boundary conditions and isolate the resonant forcing in
the bulk equations, we introduce the transformed variables
\begin{equation}
\begin{bmatrix}
\tilde{y}^{(3)}\\[2pt]
\tilde{\theta}^{(3)}
\end{bmatrix}
=
\begin{bmatrix}
y^{(3)} + \dfrac{\Delta}{\tilde{k}_1}\,\theta^{(1)}(u=1)\,u\\[8pt]
\theta^{(3)} - \dfrac{1}{6}\bigl(\theta^{(1)}\bigr)^3
\end{bmatrix}.
\label{eq:cubic_transform}
\end{equation}
By construction, (\ref{eq:cubic_transform}) removes the inhomogeneous
terms in (\ref{eq:followercubic_geom}), so that the transformed fields
satisfy homogeneous free-end conditions:
\begin{equation}
\tilde{y}^{(3)}=\tilde{\theta}^{(3)}=0\quad \mathrm{at}\quad u=0,
\label{eq:cubic_transform_clamp}
\end{equation}
\begin{equation}
\partial_u\tilde{y}^{(3)}-\nu\tilde{\theta}^{(3)}=0,\qquad
\partial_u\tilde{\theta}^{(3)}=0
\quad \mathrm{at}\quad u=1.
\label{eq:cubic_transform_free}
\end{equation}

Substituting the $O(\epsilon^3)$ kinematics
(\ref{eq:straincubic})--(\ref{eq:velocitycubic}) together with the
lower-order solutions (\ref{eq:solutionlinear}) and the $O(\epsilon^2)$
corrections into (\ref{eq:dynamicscubic}), we obtain the forced linear
system
\begin{equation}
\tilde{\gamma}_1 \partial_\tau x^{(3)} = \tilde{k}_1 \partial_u^2 x^{(3)},\quad
\partial_\tau \tilde{y}^{(3)} = \tilde{k}_2 \partial_u^2 \tilde{y}^{(3)} - \mu \partial_u \tilde{\theta}^{(3)} + f_2^{(3)},\quad
\tilde{\gamma}_3 \partial_\tau \tilde{\theta}^{(3)} = \partial_u^2 \tilde{\theta}^{(3)} + \mu (\partial_u \tilde{y}^{(3)} - \nu \tilde{\theta}^{(3)}) + f_3^{(3)},
\label{eq:cubic_forced_system}
\end{equation}
where $f_2^{(3)}$ and $f_3^{(3)}$ collect the $O(\epsilon^3)$ forcing terms
generated by (i) slow-time modulation through $\mathrm{d}A/\mathrm{d}T$,
(ii) detuning through $\Delta$, and (iii) cubic interactions involving the
$O(\epsilon)$ and $O(\epsilon^2)$ fields. Retaining only the resonant
contributions proportional to $e^{i\omega_c\tau}$, we may write
\begin{equation}
    \begin{aligned}
        f_2^{(3)} &= e^{i\omega_c \tau} \left(|A|^2A g_1 + \Delta A g_2 - Y \frac{\mathrm{d}A}{\mathrm{d}T} \right) + \mathrm{c.c.} + \mathrm{n.s.t.},\\
        f_3^{(3)} &= e^{i\omega_c \tau} \left(|A|^2A g_3 + \Delta A g_4 - \tilde{\gamma}_3\Theta \frac{\mathrm{d}A}{\mathrm{d}T} \right) + \mathrm{c.c. + \mathrm{n.s.t.}},\\
    \end{aligned}
    \label{eq:cubicforcing}
\end{equation}
where
    \begin{equation}
    \begin{aligned}
        g_1 &= i\omega_c\left(1 - \frac{\tilde{k}_2}{\tilde{k}_1} \tilde{\gamma}_1 \right) \left( 2 X_2 \Theta^*  - Y^* \Theta^2  + 2 Y |\Theta|^2\right)
        + \left( 1 - \frac{\tilde{k}_2}{\tilde{k}_1}\right) ( \Theta'^*  (P_2  + \tilde{k}_2 \Theta H)  + 2 \Theta' \,\mathrm{Re}(P_0  + \tilde{k}_2 \Theta^* H))\\
        g_2 &= \frac{i \omega_c}{\tilde{k}_1} \Theta(1) u- \left(1 - \frac{\tilde{k}_2}{\tilde{k}_1} \right) \Theta'\\
        g_3 &= \Theta^* \left( \Theta'^2 - \frac{\mu}{\tilde{k}_1} P_2 \right) + 2\Theta \left(|\Theta'|^2 - \frac{\mu}{\tilde{k}_1} \mathrm{Re}\, P_0 \right)
        -\left(1 - \frac{\tilde{k}_2}{\tilde{k}_1}\right) \left( P_2 H^* + 2 H \,\mathrm{Re} \, P_0 \right)\\
        g_4 &= \frac{\mu}{\tilde{k}_1} \left(\Theta - \Theta(1) \right) + \left(1 - \frac{\tilde{k}_2}{\tilde{k}_1} \right)H
    \end{aligned}
\end{equation}
and ``n.s.t.'' denotes non-secular terms (non-resonant contributions at
frequencies $0$, $2\omega_c$, $3\omega_c$, etc.) that do not enter the
solvability condition.

Introducing the vector $\xi^{(3)}=(x^{(3)},\tilde{y}^{(3)},\tilde{\theta}^{(3)})$ the system may
be written compactly as
\begin{equation}
\partial_\tau \xi^{(3)} = \mathcal{L}\,\xi^{(3)} + \Gamma^{-1} f^{(3)},
\label{eq:cubic_compact}
\end{equation}
together with the homogeneous boundary conditions (\ref{eq:clampingcubic}) and (\ref{eq:cubic_transform_free}), 
where $f^{(3)}=(0,f_2^{(3)},f_3^{(3)})$. In the next section, we apply the
Fredholm solvability condition associated with the adjoint eigenmode of
$\mathcal{L}$ at eigenvalue $-i\omega_c$ to eliminate resonant forcing and
obtain the Stuart-Landau equation for $A(T)$.

\subsection{Landau Coefficients}
At $\mathcal{O}(\epsilon^3)$, the perturbation fields satisfy the forced
linear system
\begin{equation}
    \partial_\tau \xi^{(3)} = \mathcal{L}\,\xi^{(3)} + \Gamma^{-1} f^{(3)},
    \label{eq:cubic_system_repeat}
\end{equation}
where $\xi^{(3)}=(x^{(3)},\tilde{y}^{(3)},\tilde{\theta}^{(3)})$ and the
operator $\mathcal{L}$ is the same non-self-adjoint linear operator
obtained at $\mathcal{O}(\epsilon)$, equipped with homogeneous boundary
conditions. The forcing $f^{(3)}$ contains contributions generated by
(i) slow-time modulation through $\mathrm{d}A/\mathrm{d}T$,
(ii) detuning from threshold through $\Delta$, and
(iii) cubic interactions among the lower-order fields.

Because the Hopf bifurcation occurs at the critical eigenvalue
$i\omega_c$, only the component of the forcing proportional to
$e^{i\omega_c\tau}$ can resonate with the neutral mode and generate
secular growth. We therefore decompose the cubic forcing as
\begin{equation}
    \Gamma^{-1} f^{(3)} = \mathcal{G}\,
    e^{i\omega_c\tau} + \mathrm{c.c.} + \mathrm{n.s.t.}.
    \label{eq:forcing_decomposition}
\end{equation}

A key feature of the follower force problem is that the linear operator
$\mathcal{L}$ is non-self-adjoint. As a result, the solvability
condition at $\mathcal{O}(\epsilon^3)$ must be imposed using the adjoint
eigenmode rather than the primal eigenmode. Moreover, because the
overdamped Cosserat-rod dynamics takes the weighted form
\begin{equation}
    \Gamma\,\partial_\tau \xi = \mathcal{M}\xi,
    \label{eq:weighted_linear_dynamics}
\end{equation}
the natural solvability condition involves the $\Gamma$--weighted inner
product
\begin{equation}
    (\Psi, \Phi)_\Gamma = \int_0^1 \mathrm{d}u \, \Psi^\dagger(u) \Gamma \Phi(u).
\end{equation}

We denote by $\xi=(0,Y,\Theta)$ the critical eigenfunction of the
primal problem at the Hopf threshold,
\begin{equation}
    \mathcal{L}\,\xi = i\omega_c\xi,
    \label{eq:primal_eigenproblem}
\end{equation}
satisfying the primal boundary conditions
\begin{equation}
Y(0) = \Theta(0)=0,\qquad Y'(1) - \nu \Theta(1) = \Theta'(1) = 0.
\end{equation}
We denote by $\Psi=(0,\Psi_Y,\Psi_\Theta)$ the corresponding adjoint
eigenfunction satisfying
\begin{equation}
    \mathcal{L}^\dagger \Psi = -i\omega_c\Psi,
    \label{eq:adjoint_eigenproblem}
\end{equation}
together with the adjoint boundary conditions
\begin{equation}
\begin{aligned}
    &\Psi_Y(0) = \Psi_\Theta(0)=0,\\
    &\tilde{k}_2 \Psi_Y'(1) - \mu \Psi_\Theta(1) = \Psi_\Theta'(1) + \mathcal{F}_c \Psi_Y(1) = 0,
\end{aligned}
\end{equation}
which we have derived in \citep{Warda2025}. We also refer to \citep{Warda2025} for the details of the semi-analytical method for computing the eigenvalues, eigenfunctions, and adjoint eigenfunctions. 

We remark that the appearance of the drag weights $\gamma_2$ and
$\gamma_3$ in (\ref{eq:inner product}) differs from the inner
product used by Schnitzer \cite{Schnitzer2025}, who considers an
inextensible Kirchhoff filament in resistive-force theory. In the
present extensible/shearable Cosserat formulation, the overdamped
dynamics is naturally expressed in the weighted form
(\ref{eq:weighted_linear_dynamics}), and the Fredholm solvability
condition is correspondingly imposed using the $\Gamma$--weighted adjoint
eigenmode.

Substituting the $\mathcal{O}(\epsilon)$ and $\mathcal{O}(\epsilon^2)$
solutions into the forcing terms in the $\mathcal{O}(\epsilon^3)$
equations yields the resonant forcing vector $\mathcal{G}$ in
(\ref{eq:forcing_decomposition}). Retaining only the contributions
proportional to $e^{i\omega_c\tau}$, we may write
\begin{equation}
    \mathcal{G}
    =
    |A|^2A \begin{bmatrix}
        0\\
        g_1\\
       \tilde{\gamma}_3^{-1}g_3
        \end{bmatrix}
        + \Delta A \begin{bmatrix}
        0\\
       g_2\\
       \tilde{\gamma}_3^{-1}g_4
    \end{bmatrix}
    - \frac{\mathrm{dA}}{\mathrm{d}T}
    \begin{bmatrix}
        0\\
        Y\\
        \Theta
    \end{bmatrix}.
\label{eq:G_vector}
\end{equation}

Since the homogeneous problem admits a solution at frequency
$e^{i\omega_c\tau}$, solvability of the inhomogeneous system requires
that the resonant forcing be orthogonal to the adjoint eigenmode under
the $\Gamma$--weighted inner product. This is the Fredholm solvability
condition,
\begin{equation}
    (\Psi,\mathcal{G})_\Gamma = 0.
    \label{eq:fredholm_condition}
\end{equation}
Substituting (\ref{eq:G_vector}) into (\ref{eq:fredholm_condition}) yields the Stuart-Landau amplitude equation governing the slow evolution of the
Hopf mode,
\begin{equation}
    \frac{\mathrm{dA}}{\mathrm{d}T}
    = \alpha |A|^2A + \beta \Delta A.
    \label{eq:stuart_landau}
\end{equation}
where the Landau coefficients $\alpha$ and $\beta$ are given by
\begin{equation}
        \alpha = \frac{\displaystyle\int_0^1 \mathrm{d}u \, \Psi_Y^* g_1 + \Psi_\Theta^* g_3}{\displaystyle\int_0^1 \mathrm{d}u \, \Psi_Y^* Y + \tilde{\gamma}_3\Psi_\Theta^* \Theta},\quad
        \beta = \frac{\displaystyle\int_0^1 \mathrm{d}u \, \Psi_Y^* g_2 + \Psi_\Theta^* g_4}{\displaystyle\int_0^1 \mathrm{d}u \, \Psi_Y^* Y + \tilde{\gamma}_3\Psi_\Theta^* \Theta}.
\end{equation}
For the numerical values of $\alpha$ and $\beta$ presented in this paper, the primal and adjoint eigenfunctions were computed numerically using the semi-analytical procedure described in \cite{Warda2025}, after which the integrals defining $\alpha$ and $\beta$ were evaluated numerically in Mathematica.

\section{Consistency Check of the Stuart-Landau Coefficient $\beta$ \label{sec:consistency}}
As a consistency check on the linear coefficient $\beta$ in the
Stuart-Landau equation, we compare it directly with the variation of
the critical eigenvalue obtained from the linear stability problem.
Near a Hopf threshold, the leading eigenvalue admits the expansion
\begin{equation}
\lambda(\tilde{\mathcal F})
=
i\omega_c
+
\left.
\frac{\mathrm{d}\lambda}{\mathrm{d}\tilde{\mathcal F}}
\right|_{\tilde{\mathcal F}_c}
(\tilde{\mathcal F}-\tilde{\mathcal F}_c)
+
\mathcal{O}\!\left((\tilde{\mathcal F}-\tilde{\mathcal F}_c)^2\right).
\end{equation}
On the other hand, using
$\tilde{\mathcal F}-\tilde{\mathcal F}_c=\epsilon^2\Delta$
and $T=\epsilon^2t$, the linearized Stuart-Landau equation,
$\mathrm{d}A/\mathrm{d}T=\beta\Delta A$, predicts the same eigenvalue variation as
\begin{equation}
\lambda(\tilde{\mathcal F})
=
i\omega_c
+
\beta(\tilde{\mathcal F}-\tilde{\mathcal F}_c)
+\cdots .
\end{equation}
Consistency therefore requires
\begin{equation}
\beta
=
\left.
\frac{\mathrm{d}\lambda}{\mathrm{d}\tilde{\mathcal F}}
\right|_{\tilde{\mathcal F}_c}.
\end{equation}

\begin{figure}[H]
\centering
\includegraphics[scale=0.55]{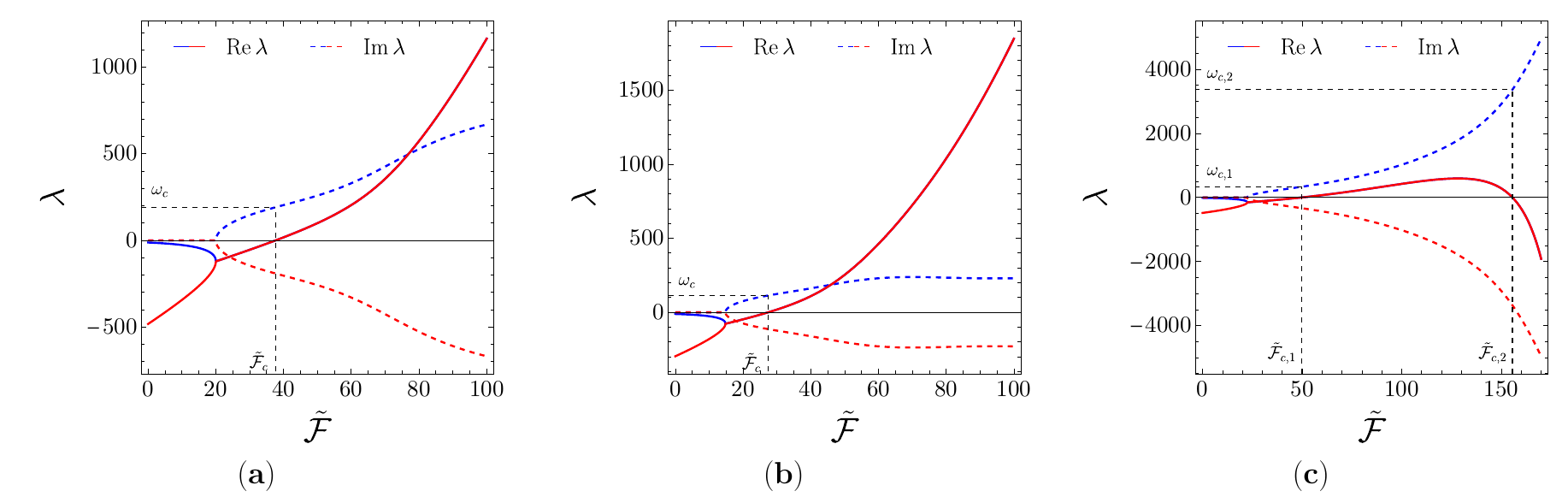}
\caption{\label{fig:beta_check}
Linear eigenvalue spectra used to verify the Stuart--Landau
coefficient $\beta$. Solid and dashed curves show, respectively, the
real and imaginary parts of the relevant eigenvalue $\lambda$ as
functions of the nondimensional follower force
$\tilde{\mathcal F}$. Panels (a)--(c) correspond to the three
constitutive parameter sets associated with the lower Hopf thresholds
listed in Table~\ref{tab:beta_check}; panel (c) additionally shows the
upper Hopf threshold of the reentrant instability window. Dashed
black lines indicate the critical follower forces and corresponding
Hopf frequencies. The slopes of the eigenvalue branches at these
thresholds are used for the consistency check reported in
Table~\ref{tab:beta_check}.
}
\end{figure}

Figure~\ref{fig:beta_check} shows the linear eigenvalue branches used
to evaluate these derivatives at the Hopf thresholds considered in
the weakly nonlinear analysis. The corresponding slopes are compared
with the independently calculated values of $\beta$ in
Table~\ref{tab:beta_check}. The derivatives were evaluated numerically from the linear stability spectrum using the centered finite-difference approximation
\begin{equation}
\left.\frac{\mathrm{d}\lambda}{\mathrm{d}\tilde{\mathcal F}}\right|_{\tilde{\mathcal F}_c}
\simeq
\frac{\lambda(\tilde{\mathcal F}_c+\delta\tilde{\mathcal F})
-\lambda(\tilde{\mathcal F}_c-\delta\tilde{\mathcal F})}
{2\delta\tilde{\mathcal F}},
\qquad
\delta\tilde{\mathcal F}=0.01.
\end{equation}
In all four cases, both the real and
imaginary parts agree to the reported numerical precision, providing
a direct consistency check on the linear term in the Stuart-Landau
equation.

\begin{table}[H]
  \centering
  \begin{ruledtabular}
    \begin{tabular}{cccccccc}
      Figure & $\tilde{k}_{1}$ & $\tilde{k}_{2}$ & $\tilde{\gamma}_{3}$ & $\tilde{\mathcal{F}}_c$& $\dfrac{\mathrm{d}\mathrm{Re} \, \lambda}{\mathrm{d} \tilde{\mathcal{F}}} \Bigg|_{\tilde{\mathcal{F}}_c}$& $\dfrac{\mathrm{d}\mathrm{Im} \, \lambda}{\mathrm{d} \tilde{\mathcal{F}}} \Bigg|_{\tilde{\mathcal{F}}_c}$ & $\beta$ (weakly nonlinear theory) \\
      \colrule
      (a) & $10^{4}$        & $10^{4}$        & $10^{-4}$ & $37.66$ & $7.35$  & $5.36$ &  $7.35 + 5.36 i$\\
      (b) & $10^{4}$        & $10^{2}$        & $10^{-2}$ & $27.43$ & $6.93$& $4.13$& $6.93+4.13 i$  \\
      (c) & $2\times 10^{2}$& $10^{4}$        & $10^{-4}$ & $49.76$ & $6.60$& $9.16$ & $6.60 + 9.16 i$ \\
      (c) & $2\times 10^{2}$        & $10^{4}$        & $10^{-4}$ & $155.60$ & $-64.59$ & $88.26$ & $-64.59 + 88.26i$ \\
    \end{tabular}
  \end{ruledtabular}
  \caption{Consistency check of the linear Stuart--Landau coefficient $\beta$.
The derivatives of the real and imaginary parts of the critical
eigenvalue with respect to the follower force are evaluated from the
linear stability calculation at each Hopf threshold
$\tilde{\mathcal F}_c$ and compared with the independently calculated
coefficient $\beta$ from Eq.~(29). In all cases,
$\left.\mathrm{d}\operatorname{Re}\lambda/\mathrm{d}\tilde{\mathcal F}\right|_
{\tilde{\mathcal F}_c}=\operatorname{Re}\beta$ and
$\left.\mathrm{d}\operatorname{Im}\lambda/\mathrm{d}\tilde{\mathcal F}\right|_
{\tilde{\mathcal F}_c}=\operatorname{Im}\beta$ to the reported
numerical precision.}\label{tab:beta_check}
\end{table}

\section{Numerical Methodology \label{sec:numerics}}

The direct numerical simulations presented in Sec. \ref{sec:simulations} were carried out
using \textit{Mathematica}'s \texttt{NDSolve} with the method of lines.
Spatial derivatives were discretized using a pseudospectral
tensor-product grid with adaptive refinement between 65 and 257
collocation points, while time integration employed the default adaptive
time-stepping scheme.

For each value of the nondimensional follower force
$\tilde{\mathcal F}$, the rod was initialized from the compressed
straight equilibrium with a small sinusoidal perturbation of amplitude
$\varepsilon=10^{-2}$ in the transverse displacement and orientation,
allowing the unstable mode to develop naturally. Simulations were
performed up to nondimensional time $\mathcal{T}=5$, which was found to be
sufficient for the transient dynamics to decay and for the oscillation
amplitude and period to become stationary for all parameter sets
considered.

The tip amplitude reported in Fig. \ref{fig:comparison} is defined as one half of the
peak-to-peak transverse displacement of the rod tip over the converged
periodic orbit. The oscillation frequency was determined from the dominant peak of the
discrete Fourier spectrum of the long-time tip displacement after
subtracting its mean value. To reduce spectral leakage, a Hann window
was applied prior to the Fourier transform. Both amplitude and frequency
were extracted using only the second half of the simulation interval,
thereby excluding the transient evolution. 

Additional simulations with longer integration times produced
indistinguishable values of the extracted amplitudes and frequencies
within the numerical accuracy reported here. As a representative convergence check, for the parameter set with $\tilde{\mathcal{F}}_c \approx 37.66$, increasing the integration time from $\mathcal{T} = 5$ to $\mathcal{T} = 10$ at $\tilde{\mathcal{F}} = 38$ changed the extracted tip amplitude by approximately $0.006 \%$, while the extracted oscillation frequency was unchanged to the reported precision.

\twocolumngrid
\bibliography{bibliography}

\begin{thebibliography}{33}%
\makeatletter
\providecommand \@ifxundefined [1]{%
 \@ifx{#1\undefined}
}%
\providecommand \@ifnum [1]{%
 \ifnum #1\expandafter \@firstoftwo
 \else \expandafter \@secondoftwo
 \fi
}%
\providecommand \@ifx [1]{%
 \ifx #1\expandafter \@firstoftwo
 \else \expandafter \@secondoftwo
 \fi
}%
\providecommand \natexlab [1]{#1}%
\providecommand \enquote  [1]{``#1''}%
\providecommand \bibnamefont  [1]{#1}%
\providecommand \bibfnamefont [1]{#1}%
\providecommand \citenamefont [1]{#1}%
\providecommand \href@noop [0]{\@secondoftwo}%
\providecommand \href [0]{\begingroup \@sanitize@url \@href}%
\providecommand \@href[1]{\@@startlink{#1}\@@href}%
\providecommand \@@href[1]{\endgroup#1\@@endlink}%
\providecommand \@sanitize@url [0]{\catcode `\\12\catcode `\$12\catcode `\&12\catcode `\#12\catcode `\^12\catcode `\_12\catcode `\%12\relax}%
\providecommand \@@startlink[1]{}%
\providecommand \@@endlink[0]{}%
\providecommand \url  [0]{\begingroup\@sanitize@url \@url }%
\providecommand \@url [1]{\endgroup\@href {#1}{\urlprefix }}%
\providecommand \urlprefix  [0]{URL }%
\providecommand \Eprint [0]{\href }%
\providecommand \doibase [0]{https://doi.org/}%
\providecommand \selectlanguage [0]{\@gobble}%
\providecommand \bibinfo  [0]{\@secondoftwo}%
\providecommand \bibfield  [0]{\@secondoftwo}%
\providecommand \translation [1]{[#1]}%
\providecommand \BibitemOpen [0]{}%
\providecommand \bibitemStop [0]{}%
\providecommand \bibitemNoStop [0]{.\EOS\space}%
\providecommand \EOS [0]{\spacefactor3000\relax}%
\providecommand \BibitemShut  [1]{\csname bibitem#1\endcsname}%
\let\auto@bib@innerbib\@empty
\bibitem [{\citenamefont {Laskar}\ \emph {et~al.}(2013)\citenamefont {Laskar}, \citenamefont {Singh}, \citenamefont {Ghose}, \citenamefont {Jayaraman}, \citenamefont {Kumar},\ and\ \citenamefont {Adhikari}}]{Laskar2013}%
  \BibitemOpen
  \bibfield  {author} {\bibinfo {author} {\bibfnamefont {A.}~\bibnamefont {Laskar}}, \bibinfo {author} {\bibfnamefont {R.}~\bibnamefont {Singh}}, \bibinfo {author} {\bibfnamefont {S.}~\bibnamefont {Ghose}}, \bibinfo {author} {\bibfnamefont {G.}~\bibnamefont {Jayaraman}}, \bibinfo {author} {\bibfnamefont {P.~B.~S.}\ \bibnamefont {Kumar}},\ and\ \bibinfo {author} {\bibfnamefont {R.}~\bibnamefont {Adhikari}},\ }\bibfield  {title} {\bibinfo {title} {Hydrodynamic instabilities provide a generic route to spontaneous biomimetic oscillations in chemomechanically active filaments},\ }\href@noop {} {\bibfield  {journal} {\bibinfo  {journal} {Sci. Rep.}\ }\textbf {\bibinfo {volume} {3}},\ \bibinfo {pages} {1964} (\bibinfo {year} {2013})}\BibitemShut {NoStop}%
\bibitem [{\citenamefont {De~Canio}\ \emph {et~al.}(2017)\citenamefont {De~Canio}, \citenamefont {Lauga},\ and\ \citenamefont {Goldstein}}]{DeCanio2017}%
  \BibitemOpen
  \bibfield  {author} {\bibinfo {author} {\bibfnamefont {G.}~\bibnamefont {De~Canio}}, \bibinfo {author} {\bibfnamefont {E.}~\bibnamefont {Lauga}},\ and\ \bibinfo {author} {\bibfnamefont {R.~E.}\ \bibnamefont {Goldstein}},\ }\bibfield  {title} {\bibinfo {title} {Spontaneous oscillations of elastic filaments induced by molecular motors},\ }\href@noop {} {\bibfield  {journal} {\bibinfo  {journal} {J. R. Soc. Interface}\ }\textbf {\bibinfo {volume} {14}},\ \bibinfo {pages} {20170491} (\bibinfo {year} {2017})}\BibitemShut {NoStop}%
\bibitem [{\citenamefont {Chirikjian}\ and\ \citenamefont {Burdick}(1994)}]{Chirikjian1994}%
  \BibitemOpen
  \bibfield  {author} {\bibinfo {author} {\bibfnamefont {G.~S.}\ \bibnamefont {Chirikjian}}\ and\ \bibinfo {author} {\bibfnamefont {J.~W.}\ \bibnamefont {Burdick}},\ }\bibfield  {title} {\bibinfo {title} {A hyper-redundant manipulator},\ }\href@noop {} {\bibfield  {journal} {\bibinfo  {journal} {IEEE Robot. Autom. Mag.}\ }\textbf {\bibinfo {volume} {1}},\ \bibinfo {pages} {22} (\bibinfo {year} {1994})}\BibitemShut {NoStop}%
\bibitem [{\citenamefont {Ranzani}\ \emph {et~al.}(2016)\citenamefont {Ranzani}, \citenamefont {Cianchetti}, \citenamefont {Gerboni}, \citenamefont {Falco},\ and\ \citenamefont {Menciassi}}]{Ranzani2016}%
  \BibitemOpen
  \bibfield  {author} {\bibinfo {author} {\bibfnamefont {T.}~\bibnamefont {Ranzani}}, \bibinfo {author} {\bibfnamefont {M.}~\bibnamefont {Cianchetti}}, \bibinfo {author} {\bibfnamefont {G.}~\bibnamefont {Gerboni}}, \bibinfo {author} {\bibfnamefont {I.~D.}\ \bibnamefont {Falco}},\ and\ \bibinfo {author} {\bibfnamefont {A.}~\bibnamefont {Menciassi}},\ }\bibfield  {title} {\bibinfo {title} {A soft modular manipulator for minimally invasive surgery: Design and characterization of a single module},\ }\href@noop {} {\bibfield  {journal} {\bibinfo  {journal} {IEEE Trans. Robot.}\ }\textbf {\bibinfo {volume} {32}},\ \bibinfo {pages} {187} (\bibinfo {year} {2016})}\BibitemShut {NoStop}%
\bibitem [{\citenamefont {Campisano}\ \emph {et~al.}(2020)\citenamefont {Campisano}, \citenamefont {Remirez}, \citenamefont {Cal\'{o}}, \citenamefont {Chandler}, \citenamefont {Obstein}, \citenamefont {Webster},\ and\ \citenamefont {Valdastri}}]{Campisano2020}%
  \BibitemOpen
  \bibfield  {author} {\bibinfo {author} {\bibfnamefont {F.}~\bibnamefont {Campisano}}, \bibinfo {author} {\bibfnamefont {A.~A.}\ \bibnamefont {Remirez}}, \bibinfo {author} {\bibfnamefont {S.}~\bibnamefont {Cal\'{o}}}, \bibinfo {author} {\bibfnamefont {J.~H.}\ \bibnamefont {Chandler}}, \bibinfo {author} {\bibfnamefont {K.~L.}\ \bibnamefont {Obstein}}, \bibinfo {author} {\bibfnamefont {R.~J.}\ \bibnamefont {Webster}},\ and\ \bibinfo {author} {\bibfnamefont {P.}~\bibnamefont {Valdastri}},\ }\bibfield  {title} {\bibinfo {title} {Online disturbance estimation for improving kinematic accuracy in continuum manipulators},\ }\href@noop {} {\bibfield  {journal} {\bibinfo  {journal} {IEEE Robot. Autom. Lett.}\ }\textbf {\bibinfo {volume} {5}},\ \bibinfo {pages} {2642} (\bibinfo {year} {2020})}\BibitemShut {NoStop}%
\bibitem [{\citenamefont {Boyer}\ \emph {et~al.}(2021)\citenamefont {Boyer}, \citenamefont {Lebastard}, \citenamefont {Candelier},\ and\ \citenamefont {Renda}}]{Boyer2021}%
  \BibitemOpen
  \bibfield  {author} {\bibinfo {author} {\bibfnamefont {F.}~\bibnamefont {Boyer}}, \bibinfo {author} {\bibfnamefont {V.}~\bibnamefont {Lebastard}}, \bibinfo {author} {\bibfnamefont {F.}~\bibnamefont {Candelier}},\ and\ \bibinfo {author} {\bibfnamefont {F.}~\bibnamefont {Renda}},\ }\bibfield  {title} {\bibinfo {title} {Dynamics of continuum and soft robots: A strain parameterization based approach},\ }\href@noop {} {\bibfield  {journal} {\bibinfo  {journal} {IEEE Trans. Robot.}\ }\textbf {\bibinfo {volume} {37}},\ \bibinfo {pages} {847} (\bibinfo {year} {2021})}\BibitemShut {NoStop}%
\bibitem [{\citenamefont {Tekinalp}\ \emph {et~al.}(2025)\citenamefont {Tekinalp}, \citenamefont {Bhosale}, \citenamefont {Cui}, \citenamefont {Chan},\ and\ \citenamefont {Gazzola}}]{Gazzola2025}%
  \BibitemOpen
  \bibfield  {author} {\bibinfo {author} {\bibfnamefont {A.}~\bibnamefont {Tekinalp}}, \bibinfo {author} {\bibfnamefont {Y.}~\bibnamefont {Bhosale}}, \bibinfo {author} {\bibfnamefont {S.}~\bibnamefont {Cui}}, \bibinfo {author} {\bibfnamefont {F.~K.}\ \bibnamefont {Chan}},\ and\ \bibinfo {author} {\bibfnamefont {M.}~\bibnamefont {Gazzola}},\ }\bibfield  {title} {\bibinfo {title} {Self-propelling, soft, and slender structures in fluids: {C}osserat rods immersed in the velocity--vorticity formulation of the incompressible {N}avier--{S}tokes equations},\ }\href@noop {} {\bibfield  {journal} {\bibinfo  {journal} {Comput. Methods Appl. Mech. Eng.}\ }\textbf {\bibinfo {volume} {440}},\ \bibinfo {pages} {117910} (\bibinfo {year} {2025})}\BibitemShut {NoStop}%
\bibitem [{\citenamefont {Laskar}\ and\ \citenamefont {Adhikari}(2017)}]{Laskar2017}%
  \BibitemOpen
  \bibfield  {author} {\bibinfo {author} {\bibfnamefont {A.}~\bibnamefont {Laskar}}\ and\ \bibinfo {author} {\bibfnamefont {R.}~\bibnamefont {Adhikari}},\ }\bibfield  {title} {\bibinfo {title} {Filament actuation by an active colloid at low {R}eynolds number},\ }\href@noop {} {\bibfield  {journal} {\bibinfo  {journal} {New J. Phys.}\ }\textbf {\bibinfo {volume} {19}},\ \bibinfo {pages} {033021} (\bibinfo {year} {2017})}\BibitemShut {NoStop}%
\bibitem [{\citenamefont {Clarke}\ \emph {et~al.}(2024)\citenamefont {Clarke}, \citenamefont {Hwang},\ and\ \citenamefont {Keaveny}}]{Clarke2024}%
  \BibitemOpen
  \bibfield  {author} {\bibinfo {author} {\bibfnamefont {B.}~\bibnamefont {Clarke}}, \bibinfo {author} {\bibfnamefont {Y.}~\bibnamefont {Hwang}},\ and\ \bibinfo {author} {\bibfnamefont {E.~E.}\ \bibnamefont {Keaveny}},\ }\bibfield  {title} {\bibinfo {title} {Bifurcations and nonlinear dynamics of the follower force model for active filaments},\ }\href@noop {} {\bibfield  {journal} {\bibinfo  {journal} {Phys. Rev. Fluids}\ }\textbf {\bibinfo {volume} {9}},\ \bibinfo {pages} {073101} (\bibinfo {year} {2024})}\BibitemShut {NoStop}%
\bibitem [{\citenamefont {Warda}\ and\ \citenamefont {Adhikari}(2026)}]{Warda2025}%
  \BibitemOpen
  \bibfield  {author} {\bibinfo {author} {\bibfnamefont {M.}~\bibnamefont {Warda}}\ and\ \bibinfo {author} {\bibfnamefont {R.}~\bibnamefont {Adhikari}},\ }\bibfield  {title} {\bibinfo {title} {Elastohydrodynamic instabilities of a soft robotic arm in a viscous fluid},\ }\href {https://doi.org/10.1103/c914-x8r2} {\bibfield  {journal} {\bibinfo  {journal} {Phys. Rev. Res.}\ }\textbf {\bibinfo {volume} {8}},\ \bibinfo {pages} {013229} (\bibinfo {year} {2026})}\BibitemShut {NoStop}%
\bibitem [{\citenamefont {Schnitzer}(2025)}]{Schnitzer2025}%
  \BibitemOpen
  \bibfield  {author} {\bibinfo {author} {\bibfnamefont {O.}~\bibnamefont {Schnitzer}},\ }\bibfield  {title} {\bibinfo {title} {Onset of spontaneous beating and whirling in the follower force model of an active filament},\ }\href@noop {} {\bibfield  {journal} {\bibinfo  {journal} {J. Fluid Mech.}\ }\textbf {\bibinfo {volume} {1007}},\ \bibinfo {pages} {A65} (\bibinfo {year} {2025})}\BibitemShut {NoStop}%
\bibitem [{\citenamefont {Beck}(1952)}]{Beck1952}%
  \BibitemOpen
  \bibfield  {author} {\bibinfo {author} {\bibfnamefont {M.}~\bibnamefont {Beck}},\ }\bibfield  {title} {\bibinfo {title} {Die knicklast des einseitig eingespannten, tangential gedr\"{u}ckten stabes},\ }\href@noop {} {\bibfield  {journal} {\bibinfo  {journal} {Z. Angew. Math. Phys.}\ }\textbf {\bibinfo {volume} {3}},\ \bibinfo {pages} {225} (\bibinfo {year} {1952})}\BibitemShut {NoStop}%
\bibitem [{\citenamefont {Bolotin}(1963)}]{Bolotin1963}%
  \BibitemOpen
  \bibfield  {author} {\bibinfo {author} {\bibfnamefont {V.~V.}\ \bibnamefont {Bolotin}},\ }\href@noop {} {\emph {\bibinfo {title} {Nonconservative {Problems} of the {Theory} of {Elastic} {Stability}}}},\ edited by\ \bibinfo {editor} {\bibfnamefont {G.}~\bibnamefont {Herrmann}}\ (\bibinfo  {publisher} {Pergamon Press},\ \bibinfo {address} {London},\ \bibinfo {year} {1963})\BibitemShut {NoStop}%
\bibitem [{\citenamefont {Ziegler}(1968)}]{Ziegler1968}%
  \BibitemOpen
  \bibfield  {author} {\bibinfo {author} {\bibfnamefont {H.}~\bibnamefont {Ziegler}},\ }\href@noop {} {\emph {\bibinfo {title} {Principles of {Structural} {Stability}}}}\ (\bibinfo  {publisher} {Blaisdell Publishing Company},\ \bibinfo {address} {Waltham, MA},\ \bibinfo {year} {1968})\BibitemShut {NoStop}%
\bibitem [{\citenamefont {Carr}\ and\ \citenamefont {Malhardeen}(1979)}]{Carr1979}%
  \BibitemOpen
  \bibfield  {author} {\bibinfo {author} {\bibfnamefont {J.}~\bibnamefont {Carr}}\ and\ \bibinfo {author} {\bibfnamefont {M.~Z.~M.}\ \bibnamefont {Malhardeen}},\ }\bibfield  {title} {\bibinfo {title} {Beck's problem},\ }\href@noop {} {\bibfield  {journal} {\bibinfo  {journal} {SIAM J. Appl. Math.}\ }\textbf {\bibinfo {volume} {37}},\ \bibinfo {pages} {261} (\bibinfo {year} {1979})}\BibitemShut {NoStop}%
\bibitem [{\citenamefont {Chen}(1987)}]{Chen1987}%
  \BibitemOpen
  \bibfield  {author} {\bibinfo {author} {\bibfnamefont {M.}~\bibnamefont {Chen}},\ }\bibfield  {title} {\bibinfo {title} {Hopf bifurcation in {B}eck's problem},\ }\href@noop {} {\bibfield  {journal} {\bibinfo  {journal} {Nonlinear Analysis: Theory, Methods \& Applications}\ }\textbf {\bibinfo {volume} {11}},\ \bibinfo {pages} {1061} (\bibinfo {year} {1987})}\BibitemShut {NoStop}%
\bibitem [{\citenamefont {Koch}\ and\ \citenamefont {Antman}(2000)}]{Koch2000}%
  \BibitemOpen
  \bibfield  {author} {\bibinfo {author} {\bibfnamefont {H.}~\bibnamefont {Koch}}\ and\ \bibinfo {author} {\bibfnamefont {S.~S.}\ \bibnamefont {Antman}},\ }\bibfield  {title} {\bibinfo {title} {Stability and {H}opf bifurcation for fully nonlinear parabolic-hyperbolic equations},\ }\href@noop {} {\bibfield  {journal} {\bibinfo  {journal} {SIAM J. Math. Anal.}\ }\textbf {\bibinfo {volume} {32}},\ \bibinfo {pages} {360} (\bibinfo {year} {2000})}\BibitemShut {NoStop}%
\bibitem [{\citenamefont {Wang}(2004)}]{Wang2004}%
  \BibitemOpen
  \bibfield  {author} {\bibinfo {author} {\bibfnamefont {Q.}~\bibnamefont {Wang}},\ }\bibfield  {title} {\bibinfo {title} {A comprehensive stability analysis of a cracked beam subjected to follower compression},\ }\href@noop {} {\bibfield  {journal} {\bibinfo  {journal} {Int. J. Solids Struct.}\ }\textbf {\bibinfo {volume} {41}},\ \bibinfo {pages} {4875} (\bibinfo {year} {2004})}\BibitemShut {NoStop}%
\bibitem [{\citenamefont {Link}\ \emph {et~al.}(2024)\citenamefont {Link}, \citenamefont {Guy}, \citenamefont {Thomases},\ and\ \citenamefont {Arratia}}]{Link2024}%
  \BibitemOpen
  \bibfield  {author} {\bibinfo {author} {\bibfnamefont {K.~G.}\ \bibnamefont {Link}}, \bibinfo {author} {\bibfnamefont {R.~D.}\ \bibnamefont {Guy}}, \bibinfo {author} {\bibfnamefont {B.}~\bibnamefont {Thomases}},\ and\ \bibinfo {author} {\bibfnamefont {P.~E.}\ \bibnamefont {Arratia}},\ }\bibfield  {title} {\bibinfo {title} {Effect of fluid elasticity on the emergence of oscillations in an active elastic filament},\ }\href@noop {} {\bibfield  {journal} {\bibinfo  {journal} {J. R. Soc. Interface}\ }\textbf {\bibinfo {volume} {21}},\ \bibinfo {pages} {20240046} (\bibinfo {year} {2024})}\BibitemShut {NoStop}%
\bibitem [{\citenamefont {Sekimoto}\ \emph {et~al.}(1995)\citenamefont {Sekimoto}, \citenamefont {Mori}, \citenamefont {Tawada},\ and\ \citenamefont {Toyoshima}}]{Sekimoto1995}%
  \BibitemOpen
  \bibfield  {author} {\bibinfo {author} {\bibfnamefont {K.}~\bibnamefont {Sekimoto}}, \bibinfo {author} {\bibfnamefont {N.}~\bibnamefont {Mori}}, \bibinfo {author} {\bibfnamefont {K.}~\bibnamefont {Tawada}},\ and\ \bibinfo {author} {\bibfnamefont {Y.~Y.}\ \bibnamefont {Toyoshima}},\ }\bibfield  {title} {\bibinfo {title} {Symmetry breaking instabilities of an in vitro biological system},\ }\href {https://doi.org/10.1103/PhysRevLett.75.172} {\bibfield  {journal} {\bibinfo  {journal} {Phys. Rev. Lett.}\ }\textbf {\bibinfo {volume} {75}},\ \bibinfo {pages} {172} (\bibinfo {year} {1995})}\BibitemShut {NoStop}%
\bibitem [{\citenamefont {Isele-Holder}\ \emph {et~al.}(2015)\citenamefont {Isele-Holder}, \citenamefont {Elgeti},\ and\ \citenamefont {Gompper}}]{Holder2015}%
  \BibitemOpen
  \bibfield  {author} {\bibinfo {author} {\bibfnamefont {R.~E.}\ \bibnamefont {Isele-Holder}}, \bibinfo {author} {\bibfnamefont {J.}~\bibnamefont {Elgeti}},\ and\ \bibinfo {author} {\bibfnamefont {G.}~\bibnamefont {Gompper}},\ }\bibfield  {title} {\bibinfo {title} {Self-propelled worm-like filaments: spontaneous spiral formation, structure, and dynamics},\ }\href {https://doi.org/10.1039/c5sm01683e} {\bibfield  {journal} {\bibinfo  {journal} {Soft Matter}\ }\textbf {\bibinfo {volume} {11}},\ \bibinfo {pages} {7181--7190} (\bibinfo {year} {2015})}\BibitemShut {NoStop}%
\bibitem [{\citenamefont {Ling}\ \emph {et~al.}(2018)\citenamefont {Ling}, \citenamefont {Guo},\ and\ \citenamefont {Kanso}}]{Ling2018}%
  \BibitemOpen
  \bibfield  {author} {\bibinfo {author} {\bibfnamefont {F.}~\bibnamefont {Ling}}, \bibinfo {author} {\bibfnamefont {H.}~\bibnamefont {Guo}},\ and\ \bibinfo {author} {\bibfnamefont {E.}~\bibnamefont {Kanso}},\ }\bibfield  {title} {\bibinfo {title} {Instability-driven oscillations of elastic microfilaments},\ }\href@noop {} {\bibfield  {journal} {\bibinfo  {journal} {J. R. Soc. Interface}\ }\textbf {\bibinfo {volume} {15}},\ \bibinfo {pages} {20180594} (\bibinfo {year} {2018})}\BibitemShut {NoStop}%
\bibitem [{\citenamefont {Fily}\ \emph {et~al.}(2020)\citenamefont {Fily}, \citenamefont {Subramanian}, \citenamefont {Schneider}, \citenamefont {Chelakkot},\ and\ \citenamefont {Gopinath}}]{Fily2020}%
  \BibitemOpen
  \bibfield  {author} {\bibinfo {author} {\bibfnamefont {Y.}~\bibnamefont {Fily}}, \bibinfo {author} {\bibfnamefont {P.}~\bibnamefont {Subramanian}}, \bibinfo {author} {\bibfnamefont {T.~M.}\ \bibnamefont {Schneider}}, \bibinfo {author} {\bibfnamefont {R.}~\bibnamefont {Chelakkot}},\ and\ \bibinfo {author} {\bibfnamefont {A.}~\bibnamefont {Gopinath}},\ }\bibfield  {title} {\bibinfo {title} {Buckling instabilities and spatio-temporal dynamics of active elastic filaments},\ }\href@noop {} {\bibfield  {journal} {\bibinfo  {journal} {J. R. Soc. Interface}\ }\textbf {\bibinfo {volume} {17}},\ \bibinfo {pages} {20190794} (\bibinfo {year} {2020})}\BibitemShut {NoStop}%
\bibitem [{\citenamefont {N{\'e}meth}\ \emph {et~al.}(2026)\citenamefont {N{\'e}meth}, \citenamefont {Warda},\ and\ \citenamefont {Adhikari}}]{Nemeth2016}%
  \BibitemOpen
  \bibfield  {author} {\bibinfo {author} {\bibfnamefont {B.}~\bibnamefont {N{\'e}meth}}, \bibinfo {author} {\bibfnamefont {M.}~\bibnamefont {Warda}},\ and\ \bibinfo {author} {\bibfnamefont {R.}~\bibnamefont {Adhikari}},\ }\href {https://arxiv.org/abs/2606.11807} {\bibinfo {title} {Shape-space dynamics and geometric pattern formation in nonreciprocal slender bodies}} (\bibinfo {year} {2026}),\ \Eprint {https://arxiv.org/abs/2606.11807} {arXiv:2606.11807 [cond-mat.soft]} \BibitemShut {NoStop}%
\bibitem [{\citenamefont {Cross}\ and\ \citenamefont {Hohenberg}(1993)}]{Cross1993}%
  \BibitemOpen
  \bibfield  {author} {\bibinfo {author} {\bibfnamefont {M.~C.}\ \bibnamefont {Cross}}\ and\ \bibinfo {author} {\bibfnamefont {P.~C.}\ \bibnamefont {Hohenberg}},\ }\bibfield  {title} {\bibinfo {title} {Pattern formation outside of equilibrium},\ }\href {https://doi.org/10.1103/RevModPhys.65.851} {\bibfield  {journal} {\bibinfo  {journal} {Rev. Mod. Phys.}\ }\textbf {\bibinfo {volume} {65}},\ \bibinfo {pages} {851} (\bibinfo {year} {1993})}\BibitemShut {NoStop}%
\bibitem [{\citenamefont {Kuznetsov}(2023)}]{Kuznetsov2023}%
  \BibitemOpen
  \bibfield  {author} {\bibinfo {author} {\bibfnamefont {Y.~A.}\ \bibnamefont {Kuznetsov}},\ }\href {https://doi.org/10.1007/978-3-031-22007-4} {\emph {\bibinfo {title} {Elements of Applied Bifurcation Theory}}}\ (\bibinfo  {publisher} {Springer International Publishing},\ \bibinfo {year} {2023})\BibitemShut {NoStop}%
\bibitem [{\citenamefont {Barkley}(2006)}]{Barkley2006}%
  \BibitemOpen
  \bibfield  {author} {\bibinfo {author} {\bibfnamefont {D.}~\bibnamefont {Barkley}},\ }\bibfield  {title} {\bibinfo {title} {Linear analysis of the cylinder wake mean flow},\ }\href {https://doi.org/10.1209/epl/i2006-10168-7} {\bibfield  {journal} {\bibinfo  {journal} {Europhysics Letters}\ }\textbf {\bibinfo {volume} {75}},\ \bibinfo {pages} {750} (\bibinfo {year} {2006})}\BibitemShut {NoStop}%
\bibitem [{\citenamefont {Tchoufag}\ \emph {et~al.}(2015)\citenamefont {Tchoufag}, \citenamefont {Fabre},\ and\ \citenamefont {Magnaudet}}]{Tchoufag2015}%
  \BibitemOpen
  \bibfield  {author} {\bibinfo {author} {\bibfnamefont {J.}~\bibnamefont {Tchoufag}}, \bibinfo {author} {\bibfnamefont {D.}~\bibnamefont {Fabre}},\ and\ \bibinfo {author} {\bibfnamefont {J.}~\bibnamefont {Magnaudet}},\ }\bibfield  {title} {\bibinfo {title} {Weakly nonlinear model with exact coefficients for the fluttering and spiraling motion of buoyancy-driven bodies},\ }\href {https://doi.org/10.1103/PhysRevLett.115.114501} {\bibfield  {journal} {\bibinfo  {journal} {Phys. Rev. Lett.}\ }\textbf {\bibinfo {volume} {115}},\ \bibinfo {pages} {114501} (\bibinfo {year} {2015})}\BibitemShut {NoStop}%
\bibitem [{\citenamefont {Antman}(2005)}]{Antman2005}%
  \BibitemOpen
  \bibfield  {author} {\bibinfo {author} {\bibfnamefont {S.~S.}\ \bibnamefont {Antman}},\ }\href@noop {} {\emph {\bibinfo {title} {Nonlinear {Problems} of {Elasticity}}}}\ (\bibinfo  {publisher} {Springer},\ \bibinfo {address} {New York},\ \bibinfo {year} {2005})\BibitemShut {NoStop}%
\bibitem [{\citenamefont {Garg}\ and\ \citenamefont {Kumar}(2023)}]{Garg2023}%
  \BibitemOpen
  \bibfield  {author} {\bibinfo {author} {\bibfnamefont {M.}~\bibnamefont {Garg}}\ and\ \bibinfo {author} {\bibfnamefont {A.}~\bibnamefont {Kumar}},\ }\bibfield  {title} {\bibinfo {title} {A slender body theory for the motion of special {C}osserat filaments in {S}tokes flow},\ }\href@noop {} {\bibfield  {journal} {\bibinfo  {journal} {Mathematics and Mechanics of Solids}\ }\textbf {\bibinfo {volume} {28}},\ \bibinfo {pages} {692} (\bibinfo {year} {2023})}\BibitemShut {NoStop}%
\bibitem [{\citenamefont {Lighthill}(1976)}]{Lighthill1976}%
  \BibitemOpen
  \bibfield  {author} {\bibinfo {author} {\bibfnamefont {J.}~\bibnamefont {Lighthill}},\ }\bibfield  {title} {\bibinfo {title} {Flagellar hydrodynamics},\ }\href {https://doi.org/10.1137/1018040} {\bibfield  {journal} {\bibinfo  {journal} {SIAM Review}\ }\textbf {\bibinfo {volume} {18}},\ \bibinfo {pages} {161} (\bibinfo {year} {1976})},\ \Eprint {https://arxiv.org/abs/https://doi.org/10.1137/1018040} {https://doi.org/10.1137/1018040} \BibitemShut {NoStop}%
\bibitem [{\citenamefont {Hinch}(1991)}]{Hinch1991}%
  \BibitemOpen
  \bibfield  {author} {\bibinfo {author} {\bibfnamefont {E.~J.}\ \bibnamefont {Hinch}},\ }\href@noop {} {\emph {\bibinfo {title} {Perturbation Methods}}},\ Cambridge Texts in Applied Mathematics\ (\bibinfo  {publisher} {Cambridge University Press},\ \bibinfo {year} {1991})\BibitemShut {NoStop}%
\bibitem [{\citenamefont {Keener}(2018)}]{Keener2018}%
  \BibitemOpen
  \bibfield  {author} {\bibinfo {author} {\bibfnamefont {J.~P.}\ \bibnamefont {Keener}},\ }\href {https://doi.org/10.1201/9780429493263} {\emph {\bibinfo {title} {Principles of Applied Mathematics: Transformation and Approximation}}}\ (\bibinfo  {publisher} {CRC Press},\ \bibinfo {year} {2018})\BibitemShut {NoStop}%
\end{thebibliography}%

\end{document}